\documentclass[%
aip,
rsi,%
amsmath,amssymb,
reprint,%
]{revtex4-1}

\usepackage{graphicx}% Include figure files
\usepackage{dcolumn}% Align table columns on decimal point
\usepackage{bm}% bold math
\usepackage{color}

\begin{document}
	
	\preprint{AIP/123-QED}

\title{Dynamics and complex formation in charged and uncharged Ficoll70 solutions}% Force line breaks with \\
%\thanks{Footnote to title of article.}

\author{Swomitra Palit}
\email{swomitra@mun.ca.}
\affiliation{Department of Physics \& Physical Oceanography, Memorial University, St. John's, NL, Canada A1B3X7}

\author{Anand Yethiraj}
\email{ayethiraj@mun.ca.}
\affiliation{Department of Physics \& Physical Oceanography, Memorial University, St. John's, NL, Canada A1B3X7}

\date{\today}% It is always \today, today,

\begin{abstract}
We apply pulsed-field-gradient NMR (PFG NMR)  technique to measure the translational diffusion for both uncharged and charged polysaccharide (Ficoll70) in water. Analysis of the data indicate that NMR signal attenuation above a certain packing fraction can be adequately fitted with a bi-exponential function. The self-diffusion measurements show also that the Ficoll70, an often-used compact, spherical polysucrose molecule, is itself non-ideal, exhibiting signs of both softness and attractive interactions in the form of a stable suspension consisting of monomers and clusters. Further, we can quantify the fraction of monomer and cluster. This work strengthens the picture of the existence of a bound water layer within and around a porous Ficoll70 particle.
\end{abstract}

\pacs{Valid PACS appear here}% PACS, the Physics and Astronomy
%Use showkeys class option if keyword
\maketitle
\section{Introduction}
\noindent A highly branched copolymer of two short building blocks, sucrose and epichlorohydrin, Ficoll70 has been widely used in studies of macromolecular crowding, and for applications in blood preservation and renal filtration due to its high hydrophobicity as well as its charge neutral globular form.~\cite{palit_combining_2017, wang_macromolecular_2012, wong2016role, Groszek_2010, Fissell2007, VenturoliF605, Dhar12102010, Wenner1999, Galan2001, lavrenko1987separation, tokuriki2004, wang_disordered_2012} This synthetic carbohydrate polymer has been used by many investigators to produce a resemblance of the high total concentrations that are encountered in the cytoplasm.~\cite{zimmerman1993} 

While some experiments found that the diffusion of Ficoll70 fits the accepted model for diffusion of hard sphere through cylindrical pores.~\cite{bohrer1984hindered, deen1981effects}, other experiments  found either that Ficoll70 was more spherical and protein-like than dextran~\cite{oliver1992determination}, or that it is more deformable than globular proteins.~\cite{axelsson2011reduced} Based on experiments in vivo, Asgeirsson {\it et al.} conjectured that Ficoll70 is sufficiently crosslinked that it cannot reptate, but is not a rigid sphere.~\cite{asgeirsson2007glomerular} Fissell and collaborators measured transport of Ficoll70 through silicon slit nanopore membranes. %with slit widths of 8–90 nm. This experiment enabled them to vary the ratio of the Ficoll70 radius to the slit width over a wide range. 
They observed that Ficoll70 molecules could penetrate the pore even when the Stokes-Einstein radius was greater than the slit width, implying deformability. They surmised Ficoll70 molecule either is not spherical, is not rigid, or exhibits a different conformation in ionic solutions.~\cite{Fissell2007}
 %Hence Ficoll70 is less well characterized in the literature than other commercial macromolecules. A distributed two-pore model was used to analyze the diffusion of Ficoll70 molecules over a porous barrier revealed the ratio of mean radius of large pore to that of small pore is equal to 2.7.~\cite{oberg2014distributed}   
%Ficoll70 lacks the strong intrachain hydrogen bonding that constrains a globular protein. It is not so compact and may be able to interpenetrate on another. Hence Ficoll70 is not a good model for globular proteins. Ficoll70 has a rather open, deformable structure comes from crowding experiments, in which the intrinsic viscosity and excluded volume have been measured as a function of solution concentration.~\cite{Wenner1999} At increasing Ficoll70 concentrations up to 2.5\%, there was an increase in the excluded volume in the solution, owing to intermolecular interactions of the Ficoll70 molecules, due to molecular crowding. However, at concentrations of 2.5\% the relative excluded volume was again reduced. This drop in excluded volume is most likely the result of molecular compression of the highly branched Ficoll70 particles. 

The most advanced analysis of Ficoll70 solution properties has been done in the renal filtration literature.~\cite{VenturoliF605,Rippe2006,Groszek_2010, OhlsonF84, Fissell2007,asgeirsson2006increased, oberg2014distributed} Fissell et al. used standard multidetector size-exclusion chromatography (SEC) on Ficoll to show that the Mark-Houwink exponents for the molecular mass dependence of the intrinsic viscosity were 0.34 (Ficoll70) and 0.36 (Ficoll400), between the value of 0 for a solid sphere and 0.5 - 0.8 for a random coil.~\cite{fissell2010size} Their result agree closely with those of Lavrenko {\it et al.}~\cite{lavrenko1986hydrodynamic} Groszek {\it et al.} used similar experiments to demonstrate that charged Ficoll70 was significantly retarded compared with uncharged Ficoll70 across the rat glomerular filtration barrier.~\cite{Groszek_2010} Georgalis {\it et al.} found two different sizes of particles in Ficoll70 by means of light scattering experiments.~\cite{georgalis2012light} %Measurements of intrinsic viscosity of Ficoll70  solutions as a function of volume fraction~\cite{Wenner1999} also strongly suggest that the Ficoll70 particles have a much more open structure than has been previously commonly conceived.

\begin{table*}
	\caption{Comparison of the zeta potential for charged and uncharged Ficoll70}
	\begin{ruledtabular}
		\begin{tabular}{cccc}
			Species & Zeta Potential & Mobility & Conductivity \\
			& (mV) & ($\mathrm \mu$m cm V/s) & (mS/cm) \\
			\hline \\
			Charged Ficoll70 & -27 $\pm$ 4 & - 1.4 $\pm$ 0.2 & 0.1 $\pm$ 0.02 \\
			(without salt) & & & \\
			Charged Ficoll70 & -29 $\pm$ 2 & -1.3 $\pm$ 0.4 & 1.1 $\pm$ 0.01\\
			(salt added) & & & \\
			Uncharged Ficoll70 & -5.2 $\pm$ 0.2 & - 0.4 $\pm$ 0.02 & 0.04 $\pm$ 0.01 \\
		\end{tabular}
	\end{ruledtabular}
\end{table*}

\begin{figure}[!bht]
	\centering
	\includegraphics[height=65mm]{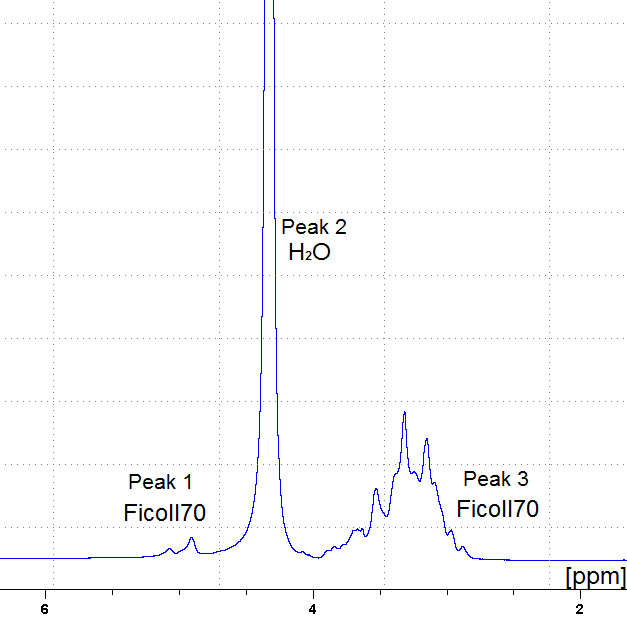}%\end{subfigure} 
	\caption{1D $^{1}$H-NMR spectrum for Ficoll70 /$\mathrm{H_{2}O}$ sample at a sample temperature 298 K.}
	\label{OneDNMR}
\end{figure}

In this study, we employ pulsed-field-gradient (PFG) NMR to monitor the self-diffusivities of uncharged and charged Ficoll70 in deionized water. Because of the spectral selectivity of NMR, we can simultaneously (see Figure~\ref{OneDNMR}) obtain signal from both the Ficoll70 and water species. 
In a companion work, we focus on polymer structure and dynamics~\cite{sp2} in the presence of Ficoll70 crowder. Ficoll is an often-used crowder. In the understanding of macromolecular crowding, it is important to understand well the properties of the crowder. In this work, we examine the properties of both charged and uncharged Ficoll70 for evidence of cluster formation in equilibrium, a phenomenon, distinct from bulk phase separation, that has been identified in colloids and proteins where short-ranged attractions coexist with longer-ranged (typically electrostatic) repulsive interactions.~\cite{groenewold_anomalously_2001, stradner_equilibrium_2004, barhoum_clusters_2010, porcar_formation_2010, barhoum_cluster_2013, sweatman_cluster_2014}
%, and find that a) mixtures of charged and un-charged Ficoll can be used to mimic PEG dynamics in cell lysates. and b) it is not a hard-sphere crowder.

\section{Materials and Methods}\label{sec:material}
Ficoll\textregistered PM 70 (referred to as Ficoll70 in the text) with average molecular weight of 70000 (mean radius 4.5-5.5 nm~\cite{sorensson1998glomerular, georgalis2012light, Wenner1999, luby1987hindered}) was purchased from Sigma Aldrich and used without further purification. Charged Ficoll70
(Ficoll CM 70) was a carboxymethylated derivative of Ficoll PM70, made as described in reference.~\cite{Groszek_2010} It was a gift from Dr. William H. Fissell, and was used as received after having been neutralized and dialyzed against distilled water for 4 days. Experimental volume fractions of Ficoll70 were calculated using the partial specific volume of Ficoll70, which is 0.67 $\mathrm {cm^3}$/g.~\cite{lavrenko1987separation} 

For sample preparation, the desired volume fraction of Ficoll70 was dissolved in deionized H$_2$O. 
For charged Ficoll70 solutions, the conductivity was controlled, using KCl, to a value of $\approx 1$ mS/cm (see Table 1) in order to ensure a consistent Debye-H\"{u}ckel screening length for all samples.
The solution was stirred for 10 hours. Samples were then transferred to 5 mm outer diameter NMR tubes.
\begin{figure*}[!bht]
	\centering
	\includegraphics[height=60mm]{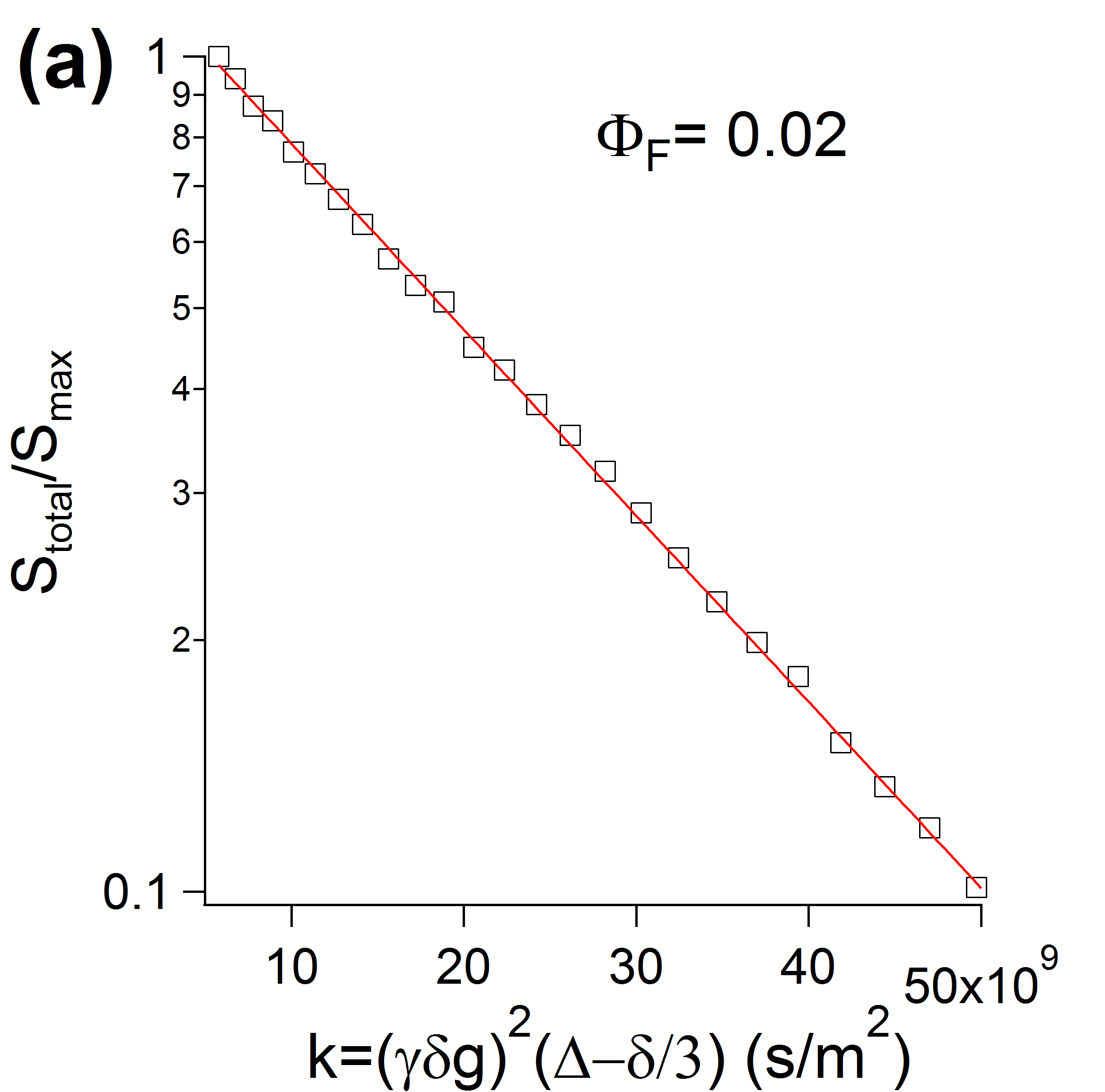}
	\includegraphics[height=60mm]{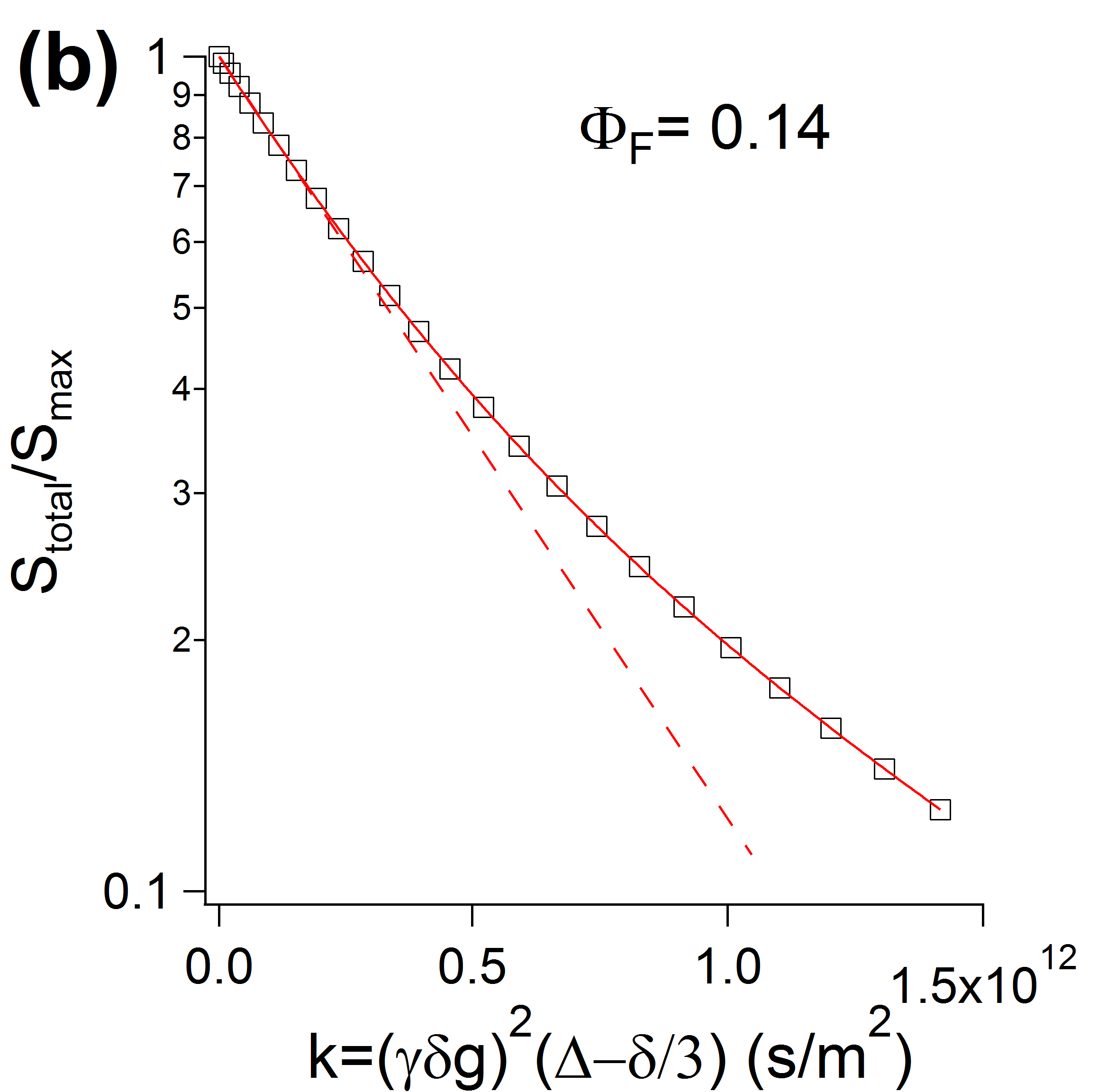}
	\caption{ (a) The attenuation of the signal $\mathrm S_{\mathrm {total}}/\mathrm S_{\mathrm {max}}$ on a log scale versus the gradient strength parameter $\mathrm {k}=\mathrm{(\gamma \delta g)^{2} (\Delta-\delta/3)}$ for an aqueous solution Ficoll70 is mono-exponential at low $\mathrm {\Phi_F}$ for both uncharged and charged Ficoll70 solutions. Signal attenuation for charged Ficoll70 solution at $\mathrm{\Phi_F=0.02}$ exhibits simple mono-exponential behaviour. (b) For $\mathrm{\Phi_F > 0.05}$ (0.10) for uncharged (charged) Ficoll70, the signal attenuation is not mono-exponential. As an example, at $\mathrm{\Phi_F = 0.14}$, the signal attenuation is well-fit to a bi-exponential form (solid line). The dashed line represents mono-exponential fit to the first few points.}
	\label{fig:diffcurve}
\end{figure*}

%\begin{figure}[!bht]
%	\centering
%	\includegraphics[height=80mm]{signalatten}%\end{subfigure} 
	%\includegraphics[height=60mm]{signalatten}%\end{subfigure} %
%	\caption{The PFG NMR signal attenuation for Ficoll70/$\mathrm {H_{2}O}$ system. Here the gradient pulse strength g was increased in a linear sequence of 32 steps up to 200 G/cm}
%	\label{Ficollattenuation}
%\end{figure}

\begin{figure}[!htp]
	\centering
	%	\begin{subfigure} {.4\textwidth} \centering
	\includegraphics[width=60mm]{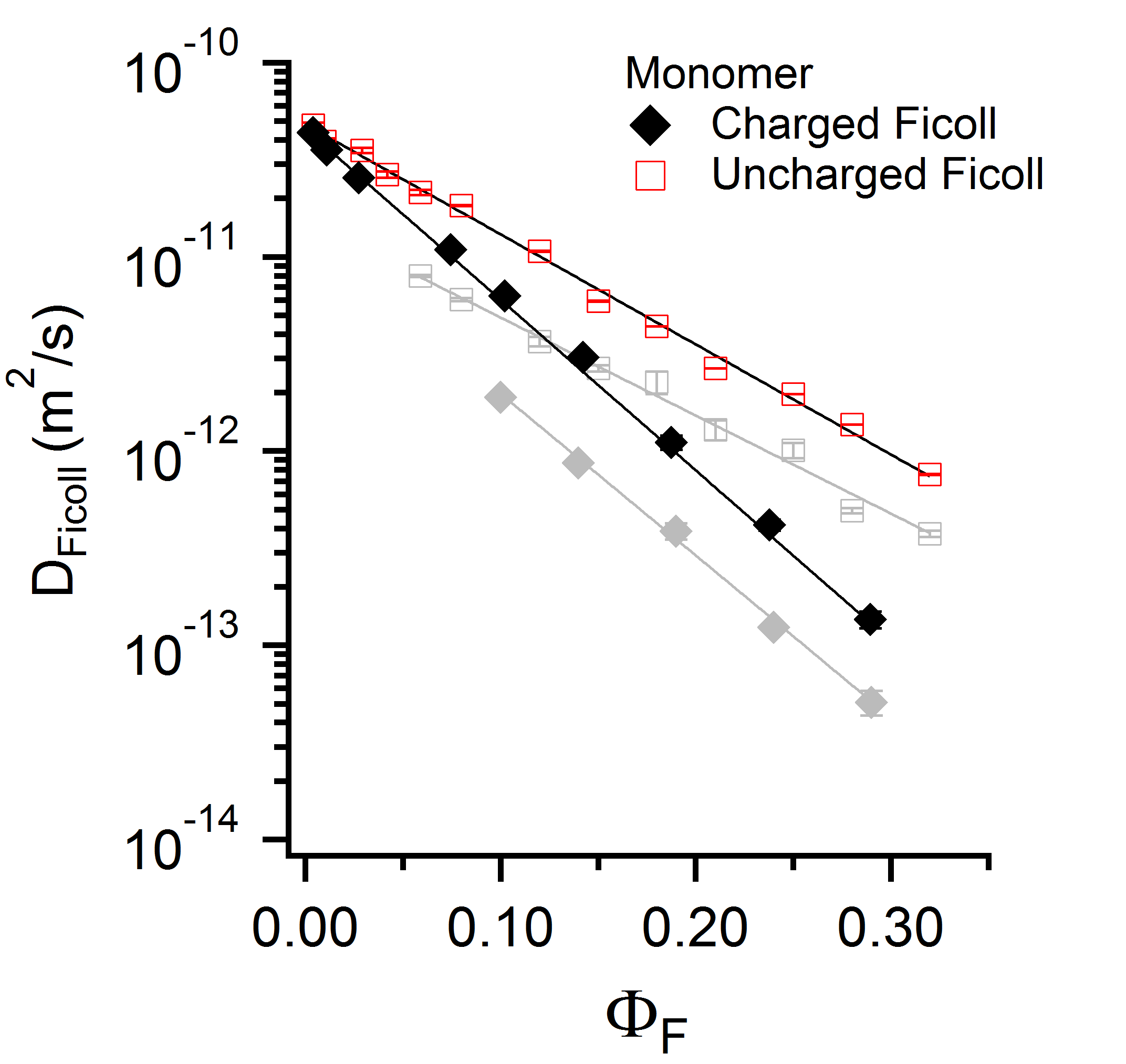} %\end{subfigure} %
	%	\begin{subfigure} {.4\textwidth} \centering
	\includegraphics[width=60mm]{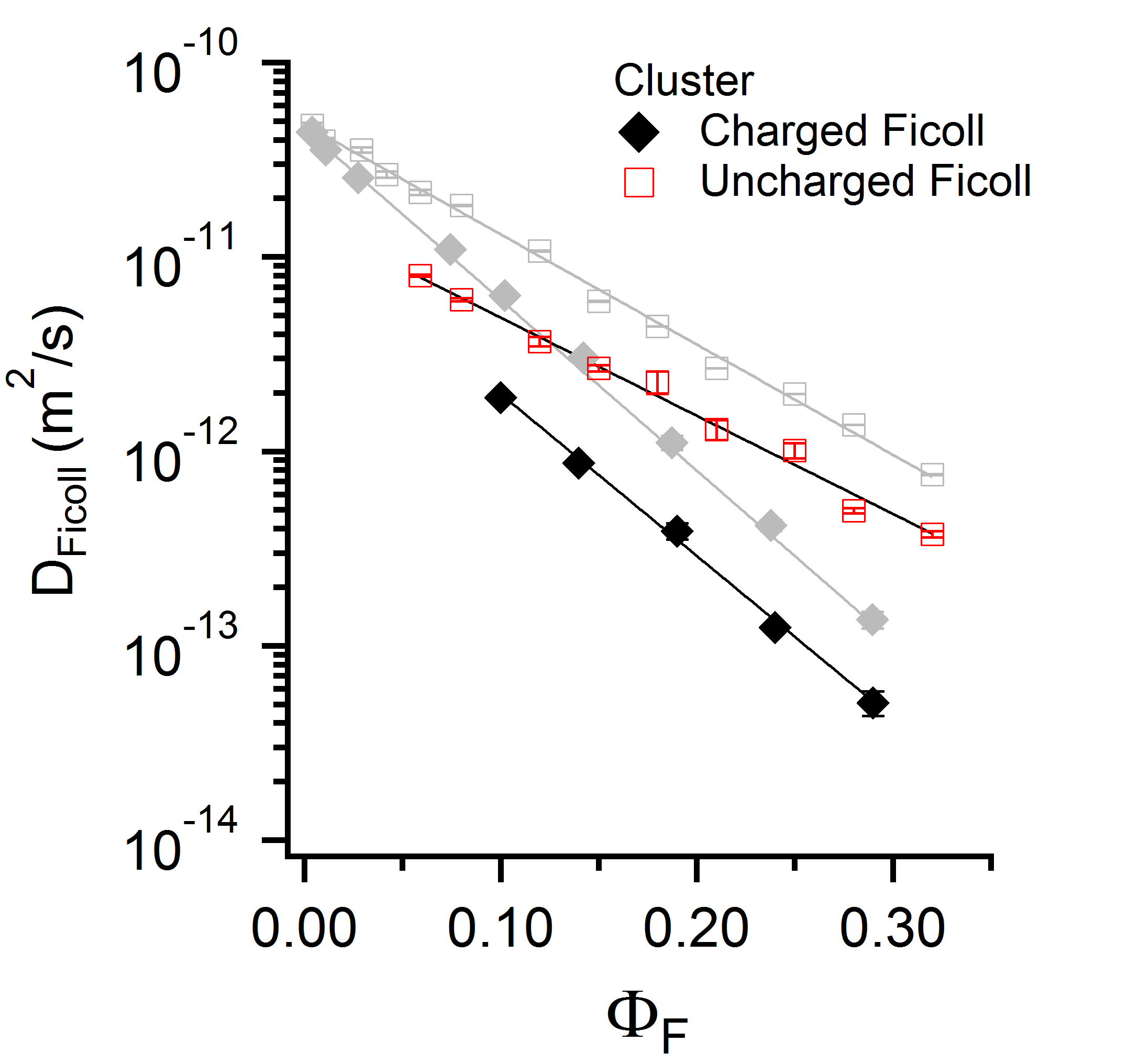}
	\caption{{\bf Ficoll70 forms clusters:} Biexponential signal attenuation indicates emergence of a cluster state above $\mathrm {\Phi_F= 0.05}$ (uncharged) and $\mathrm{\Phi_F= 0.1}$ (charged). (a) Ficoll70 monomer diffusion coefficient as a  function of $\mathrm {\Phi_F}$ and (b) Ficoll70 cluster diffusion coefficient as a function of $\mathrm {\Phi_F}$. In (a) and (b) cluster and monomer diffusion results are shown in gray to aid comparison.}
	\label{fig:DFicollMonomerAggregate}
\end{figure}

\subsection{PFG NMR}
The one-dimensional  1D proton NMR spectrum has been observed for different species in all samples at a resonance frequency of 600 MHz on a Bruker Avance II spectrometer. Figure~\ref{OneDNMR}  shows well-separated peak regions related to this system. Peak 1 and Peak 3 are the Ficoll70 peaks whereas Peak 2 is for $\mathrm {H_2O}$ molecules in solution. 
All NMR experiments were performed at T =298K. The self-diffusion measurements were carried out in a diffusion probe  Diff 30 and with maximum field gradient  1800 G/cm (18 T/m). Diffusion was measured with a pulsed-field-gradient stimulated echo sequence with trapezoidal gradient pulses.~\cite{Price1997} The diffusion coefficient of a molecule in aqueous solution is obtained from the attenuation of the signal according to the equation
\begin{equation}
\mathrm{S(k) = S(0)\exp(-D k)}, 
\end{equation}
where $\mathrm{S(k)}$ is the intensity of the signal in the presence of field gradient pulse, $\mathrm{S(0)}$ is the intensity of the signal in the absence of field gradient pulse, $\mathrm{k=(\gamma\delta\Delta)^{2}(\Delta-\delta/3)}$, $\mathrm{\gamma=\gamma^{H}=2.657 \times 10^{8}}$ $\mathrm{T^{-1}.s^{-1}}$ 
is the proton gyromagnetic ratio, $\mathrm{\delta}$= 2 ms is the duration of field gradient pulse, $\mathrm {\Delta}$= 100 ms is the time period between two field gradient pulses, and g is the amplitude of field gradient pulse.

\subsection{Zeta Potential}
The Zeta potential ($\mathrm \zeta$) and electrophoretic mobility of Ficoll70 solutions, shown in Table 1, were measured by a  Zetasizer Nano Z system (Malvern Instruments Ltd, Malvern, United Kingdom). 
The dimensionless Zeta potential $\mathrm {\Psi = \zeta e/k_BT = 1.1 \pm 0.2}$ and $0.21 \pm 0.02$ for charged and uncharged Ficoll70 respectively. The solutions of charged Ficoll70 were all prepared with added salt in order to keep the conductivity at 1 mS/cm, resulting in a Debye-H\"{u}ckel screening length $\mathrm {\mathrm \kappa^{-1} = 3.2 \pm 0.5}$ nm. This corresponds to a $\mathrm {\kappa R_c \sim 1.4}$. Given the value of the dimensionless Zeta potential $\mathrm \Psi $ and $\mathrm {\kappa R_c}$, i.e., both of order unity, electrostatics should clearly be important, but not overwhelmingly so.

\subsection{Bulk Viscosity Measurement}
Experiments were performed on an Anton Paar Physica MCR 301 rheometer, where the cone-plate measuring system was used to extract the flow curves. The cone-plate geometry has a diameter of 50 mm and cone angle of $0.5^0$. The flow curves experiments were carried out with shear rate varying from 0.001 to 150 s$^{-1}$. For all samples reported in this work, viscosity remains constant as the shear rate is varied. 

\begin{figure*}[!htb]
	\centering
	%	\begin{subfigure} {.4\textwidth} \centering
	%	\includegraphics[width=70mm]{Uncharged_Charged_Monomer_Diffusion_new_.png}%\end{subfigure} %
	%	\begin{subfigure} {.4\textwidth} \centering
	%	\includegraphics[width=70mm]{Uncharged_Charged_Aggregate_Diffusion_New_.png}
	\includegraphics[width=60mm]{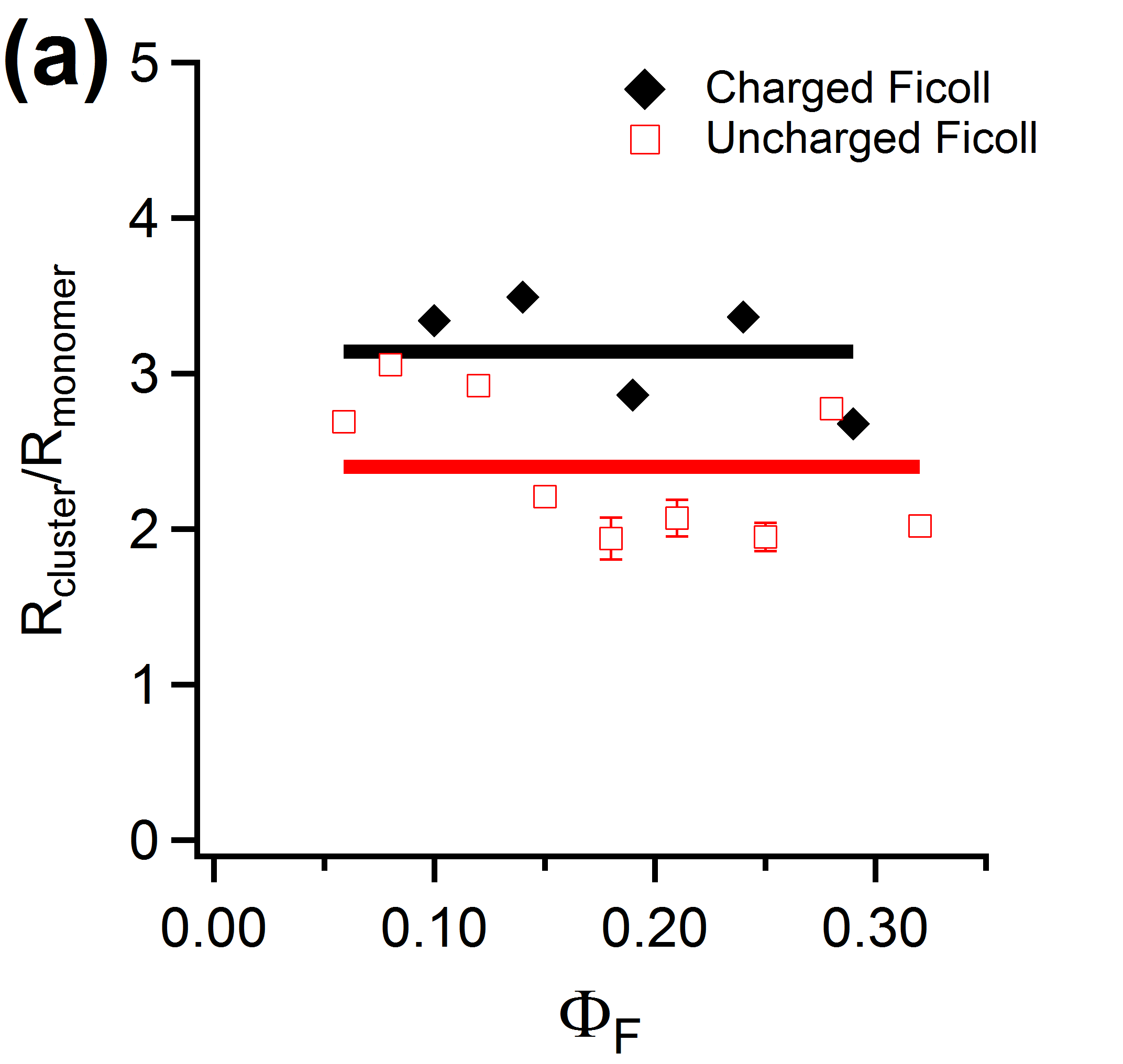}
	\includegraphics[width=60mm]{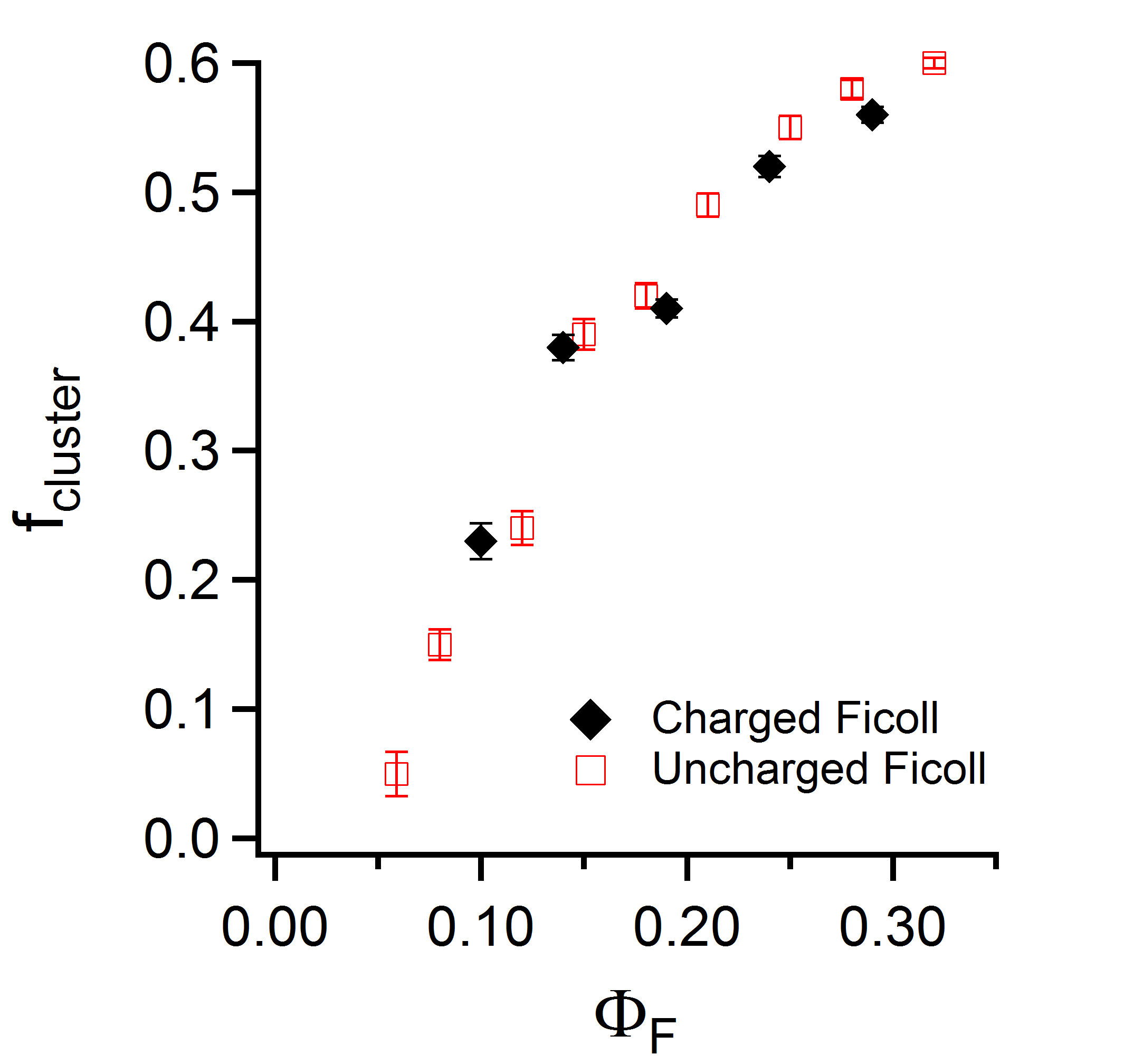}
	\caption{{\bf Structure of Ficoll70 via diffusion:} (a) Ficoll70 cluster to monomer size ratio and
		(b) fraction of Ficoll70 cluster ($\mathrm {f_{cluster}}$) as a function of $\mathrm {\Phi_F}$ for both charged and uncharged Ficoll70. }
	\label{fig:FicollFraction}
\end{figure*}

\section{Diffusion Model}
The PFG NMR signal attenuation of Ficoll70 shows a monoexponential decay with the gradient strength parameter at low packing fraction ($\mathrm {\Phi_F} < 0.05$ (uncharged) and $\mathrm {\Phi_F} < 0.1$ (charged)). This implies either that it is a single component system or that there are multiple components (e.g. a monomer and cluster) that exchanges very rapidly between monomer and aggregate on the timescale of the NMR experiment.~\cite{barhoum_diffusion_2016}  Given the larger size of Ficoll70, the diffusion time of the monomer $\sim 1 \mu$s; thus residence times of the Ficoll70 molecule within clusters will be a few micro-seconds or longer.  Hence the fact that the signal attenuation associated with the Ficoll70 peak exhibits monoexponential behaviour (Figure~\ref{fig:diffcurve}(a)) at low packing fractions suggests that the exchange between Ficoll70 clusters and monomers must be very rapid on the NMR time scale.

On the other hand, if the molecular exchange between monomer and cluster is very slow, one expects the total Ficoll70 signal to be given by 
\begin{eqnarray}
\mathrm S_{\mathrm{total}} &=& \mathrm S_{\mathrm{monomer}} + \mathrm S_{\mathrm{cluster}} \nonumber \\
&=& \mathrm S_{\mathrm{0,monomer}} \exp(-\mathrm D_{\mathrm{monomer}}{\mathrm k}) \nonumber \\ 
&+ & \mathrm S_{\mathrm{0,cluster}} \exp(-\mathrm D_{\mathrm{cluster}}{\mathrm k}) 
\label{diffnoexchange}
\end{eqnarray}

which is bi-exponential in nature (Figure~\ref{fig:diffcurve}(b)). A generalization to multi-exponential behaviour may be made for macromolecules existing in more than two species: $\mathrm S_{\mathrm{total}} = \sum_i \mathrm S_{0,\mathrm{i}} \exp(-\mathrm D_{\mathrm{i}}{\mathrm k})$. For two species, Equation~\ref{diffnoexchange} may be written in the form $\mathrm S_{\mathrm{total}}/\mathrm S_{\mathrm{max}} = \mathrm f \exp(-\mathrm D_{\mathrm{1}}{\mathrm k}) + (1 - \mathrm f) \exp(-\mathrm D_{\mathrm{2}}{\mathrm k})$, where $\mathrm f =  \mathrm S_{0,\mathrm{1}}/(\mathrm S_{0,\mathrm{1}} + \mathrm S_{0,\mathrm{2}})$.

\section{Results}
The spectral selectivity of PFG NMR allows us to simultaneously obtain diffusion coefficients of water and Ficoll70. We can thus obtain, not only Ficoll70 dynamics, but also the information about the interaction of water with the crowder.

%\subsection{Ficoll is not a hard sphere}\label{sec:hydro}
\subsection{Ficoll70 forms clusters}
The crowder diffusion coefficient is obtained in pure Ficoll70 aqueous solutions. The key observation is that the PFG NMR signal attenuation is not mono-exponential when $\mathrm{\Phi_F}$ is greater than a threshold value: 0.05 (0.10) for uncharged (charged) Ficoll70. When there are two species with the same chemical signatures, and when there is slow exchange (or no exchange) between the species, one obtains bi-exponential signal attenuations in a PFG NMR experiment (Figure~\ref{fig:diffcurve}(b)). Our observations thus indicate the co-existence of (fast diffusing) monomers and (slow diffusing) clusters of Ficoll70.

We plot the diffusion coefficients for charged and uncharged crowder, and for monomer (Figure~\ref{fig:DFicollMonomerAggregate}(a)) and for cluster 
(Figure~\ref{fig:DFicollMonomerAggregate}(b)), as a function of $\mathrm{\Phi_F}$. Every $\mathrm D$ dependence on $\mathrm {\Phi_F}$ is exponential! 
As discussed in earlier~\cite{palit_combining_2017} and companion~\cite{sp2} works, the work of Rosenfeld~\cite{rosenfeld_relation_1977} and Dzugutov~\cite{dzugutov_universal_1996}
connected structural properties of atomic fluids to their diffusion coefficients. 
Both studies have proposed an exponential relationship between atomic diffusion and the excess entropy $\mathrm {S_2/k_B}$ (in the 2-particle approximation); moreover, recent 2D simulations and colloids experiments~\cite{thorneywork_effect_2015} show that $\mathrm {S_2/k_B}$ is proportional to the colloid packing fraction for packing fractions less than 0.4. The same connection would hold in colloidal suspensions if hydrodynamics is not important in the long-time limit.

\begin{figure}[!htb]
	\centering
	\includegraphics[width=80mm]{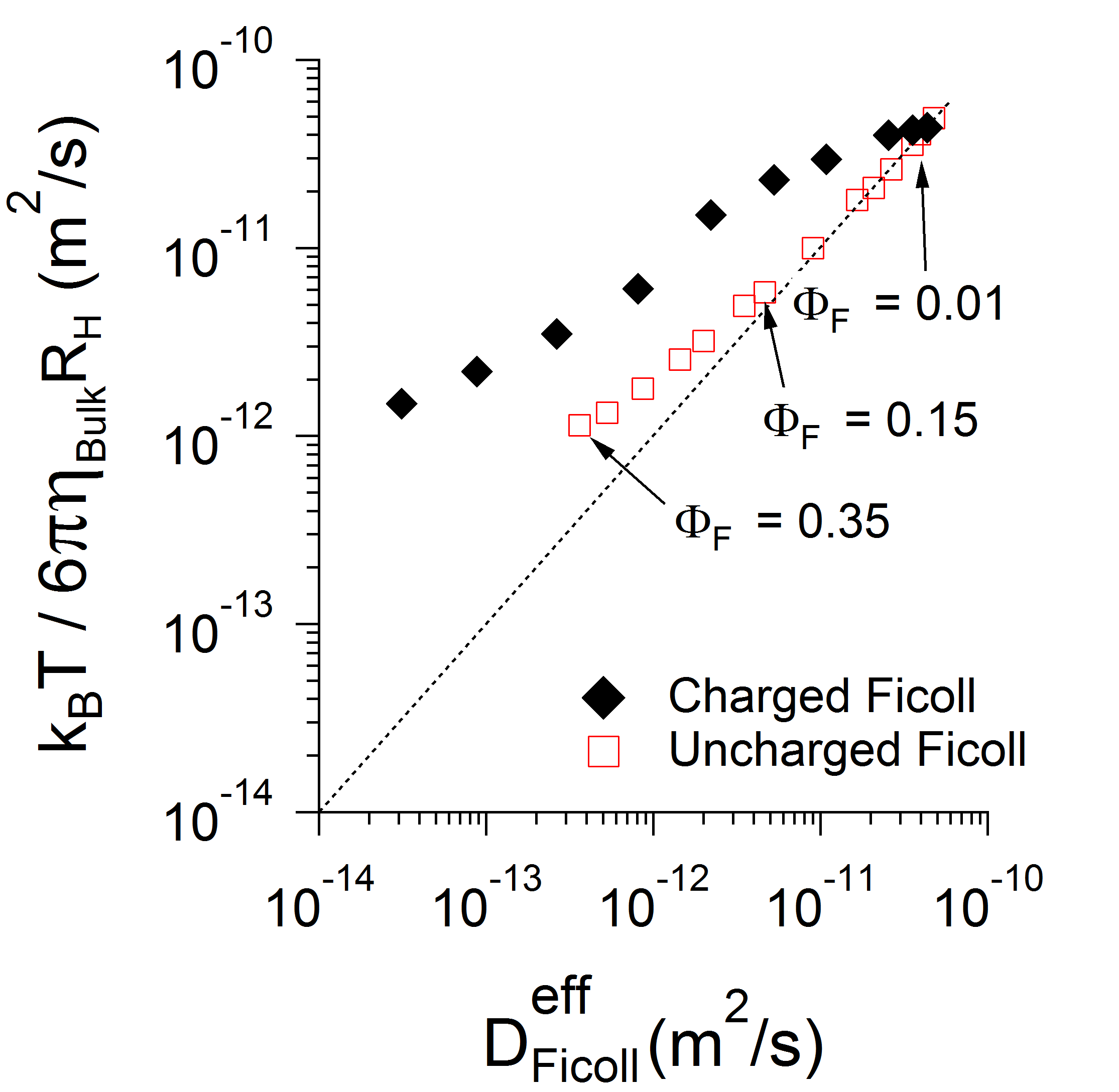}
	\caption{{\bf Effective diffusion coefficient of Ficoll70:} Comparison of a self-diffusivity $\mathrm{k_B T / 6 \pi \eta_{Bulk} R_H}$, calculated from the bulk Ficoll70 viscosity $\mathrm {\eta_{Bulk}}$ and the hydrodynamic radius of Ficoll70 monomer $\mathrm{R_H = 4.6}$ nm, as a function of the measured effective diffusion coefficient $\mathrm{D^{eff}}$, shows agreement with Stokes-Einstein behaviour (dashed line) upto $\mathrm{\Phi_F}= 0.15$ for uncharged Ficoll70, while for charged Ficoll70 there is significant deviation for much smaller $\mathrm{\Phi_F}$.}
	
	% demonstrates that the compact Ficoll70 monomer is slowed down significantly more, at high $\Phi_F$, than the polymer chain: by a factor of 10 and 100 for uncharged and charged Ficoll70.
	\label{fig:viscodiffficoll}
\end{figure}

From the ratio of monomer and cluster diffusion coefficient at each packing fraction $\mathrm {\Phi_F}$, we can make a crude estimate of the size of the cluster. 
This is estimated assuming $\mathrm{R_{\mathrm{cluster}}/R_{\mathrm{monomer}} = D_{\mathrm{monomer}}/D_{\mathrm{cluster}}}$, i.e., that the local micro-environment for monomer and cluster are identical. For uncharged Ficoll70, Georgalis et. al. measured the value of $\mathrm {D_{\mathrm{monomer}}/D_{\mathrm{cluster}}= 2.37}$.~\cite{georgalis2012light} 
The cluster-to-monomer size ratio (Figure~\ref{fig:FicollFraction}(a)) shows no clear dependence on $\mathrm{\Phi_F}$, but appears somewhat larger for charged Ficoll70 than for uncharged Ficoll70.  
%Indeed, we will see in the next section that the local micro-viscosities can be very different from the bulk viscosity, so the values for $R_{\mathrm{cluster}}/R_{\mathrm{monomer}}$ are an upper bound.
The fraction of clusters (shown in Figure~\ref{fig:FicollFraction}(b)) increases from 5\% at onset of clustering to $\sim 60\%$ in the crowding regime: in fact, this fraction is very similar for charged and uncharged crowder.

One can use the measured monomer and cluster self-diffusivities to calculate an effective diffusion coefficient $\mathrm{D_{eff}}$
\begin{equation}
\label{AverageWeightDiffusion}
\mathrm {D^{eff}_{Ficoll} = f_{cluster} D_{cluster} + (1 - f_{cluster}) D_{monomer}}.
\end{equation}
This diffusivity may be compared to its bulk analog from the measured bulk Ficoll70 viscosity $\mathrm {\eta_{Bulk}}$ and the hydrodynamic radius of Ficoll70 monomer $\mathrm{R_H = 4.6}$ nm using a Stokes-Einstein form $\mathrm{k_B T / (6 \pi \eta_{Bulk} R_H)}$. A slope of 1 in the plot of $\mathrm{k_B T / (6 \pi \eta_{Bulk} R_H)}$ versus $\mathrm{D_{eff}}$ would imply agreement with Stokes-Einstein behaviour (dashed line). As can be seen, there is agreement upto $\mathrm{\Phi_F = 0.15}$ for uncharged Ficoll70, while for charged Ficoll70 there is significant deviation for much smaller $\mathrm{\Phi_F}$.
Even for uncharged Ficoll70 solutions, there is significant deviation for $\mathrm{\Phi_F > 0.15}$.

%\ay{Hence we obtain $\mathrm {D^{eff}_{Ficoll}}$ from the fraction of cluster $\mathrm{f_{cluster}}$ and the fraction of monomer ($1-\mathrm{f_{cluster}}$), for a range of $\mathrm{\Phi_F}$, and the diffusivities $\mathrm{D_{cluster}}$ and $\mathrm{D_{monomer}}$ of both cluster and monomer species.}

%\ay{According to the Stokes-Einstein relation, the diffusion coefficient of a spherical molecule depends inversely on the viscosity of the surrounding medium. Here we independently measure the bulk viscosity of Ficoll70 and effective diffusion coefficient using Equation~\ref{AverageWeightDiffusion}. The results are shown in Figure~\ref{fig:viscodiffficoll}(b). These results show that inverse of the bulk viscosity depends nonlinearly on the diffusion coefficient of the Ficoll70. This is expected as Stokes-Einstein relation derived for a structureless, continuum viscous fluid. }

\subsection{Ficoll is porous}
\begin{figure}[!htb]
	\centering
	%	\begin{subfigure} {.4\textwidth} \centering
	%	\includegraphics[width=70mm]{Uncharged_Charged_Monomer_Diffusion_new_.png}%\end{subfigure} %
	%	\begin{subfigure} {.4\textwidth} \centering
	%	\includegraphics[width=70mm]{Uncharged_Charged_Aggregate_Diffusion_New_.png}
	%	\includegraphics[width=60mm]{R_Aggregate_new_.png}
	%	\includegraphics[width=60mm]{Fraction_Aggregate_new_.png}
	\includegraphics[width=80mm]{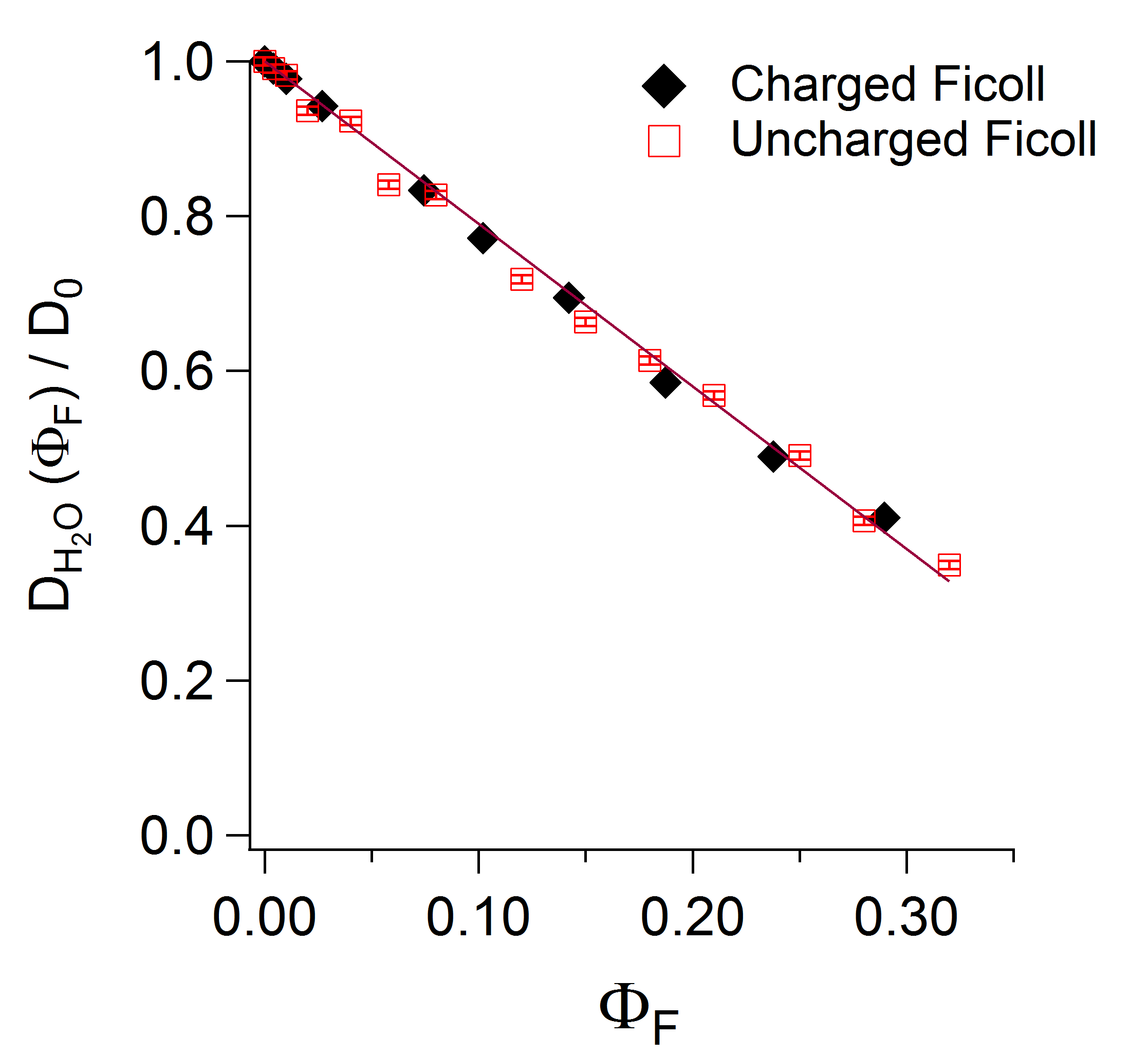}
	\caption{{\bf Ficoll70 is porous:} Linear decrease in water diffusion coefficient with increasing $\mathrm{\Phi_F}$ indicates a fraction of surface-associated water.
	}
	\label{fig:waterDiffusion}
\end{figure}

Another interesting aspect is the water diffusion coefficient. The similarity of the water diffusion for charged and uncharged Ficoll70 in Figure~\ref{fig:waterDiffusion} is reassuring, as it indicates that the physical structure of the polysucrose is unchanged by the charge. A linear decrease in water diffusion coefficient is observed with increasing $\mathrm {\Phi_F}$, which indicates a fraction of surface-associated water $\mathrm {f_{surface} \approx D_{\mathrm{H_2O}} (\mathrm{\Phi_F})/D_0}$, shown in Figure~\ref{fig:waterDiffusion}, that scales with crowder packing fraction. For solid, spherical colloids the fraction of ``bound water'' would be expected to scale with the total surface area of the particles, not the volume. The high degree of linearity in Figure~\ref{fig:waterDiffusion} with a fit to $\mathrm {\mathrm {D_{\mathrm{H_2O}}/D_0 = 1 - \beta_1 \Phi_F}}$, with $\mathrm {\mathrm \beta_1 = 2.10 \pm 0.03}$ implies that the Ficoll70 is a porous particle and contains a significant amount of water inside its polysucrose matrix.

At $\mathrm{\Phi_F} = 0.3$, as much as 60\% of the water is surface associated, suggesting that Ficoll70 is highly porous. 
The porous nature of Ficoll70 is not surprising, in hindsight, but we believe that it has not been adequately recognized in the crowding literature, apart from clear indications that Ficoll70 is not a rigid sphere~\cite{Fissell2007,VenturoliF605}, as well as the practical knowledge about the lack of overall stability of Ficoll70 solutions above $\mathrm {\Phi_F = 0.35\%}$ . It should be noted that this bound water is likely not available to the polymer, and should be accounted for in any free-volume calculations.

\subsection{The effect of PEG polymer on Ficoll diffusion}
\begin{figure}[!htb]
	\centering
	\includegraphics[width=80mm]{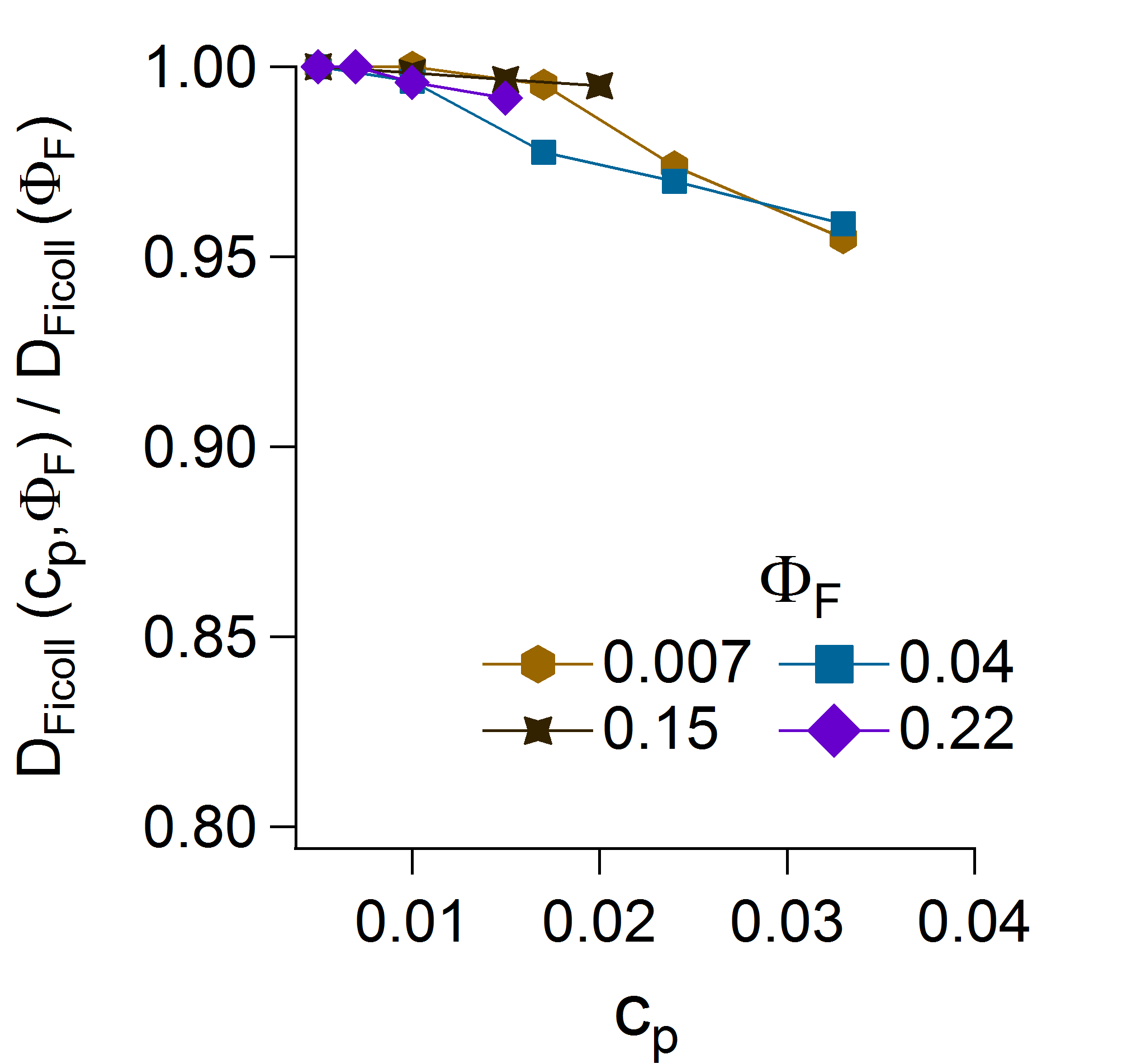}
	\caption{{\bf Effect of polymeric PEG macromolecules on the dynamics of Ficoll70:} The crowder (Ficoll70) diffusion coefficient is obtained in the presence of small amounts of PEG ($\mathrm {0.003 - 0.03 g/cm^3}$). Here PEG has no effect on Ficoll70 dynamics, except for a weak $\mathrm{c_p}$ dependence at the lowest $\mathrm{\Phi_F}$.}

	% demonstrates that the compact Ficoll70 monomer is slowed down significantly more, at high $\Phi_F$, than the polymer chain: by a factor of 10 and 100 for uncharged and charged Ficoll70.
	\label{fig:diffficoll}
\end{figure}

Figure~\ref{fig:diffficoll} shows plots of the diffusion coefficient of Ficoll70 for different packing fraction $\mathrm{\Phi_F}$, in the presence of small amounts of a polymeric macromolecule, polyethyelene glycol (PEG), in the concentration range $\mathrm{0.003 - 0.03 g/cm^3)}$. PEG is the macromolecule used in the companion work~\cite{sp2}. A weak effect on Ficoll70 dynamics has been observed at the lowest $\mathrm {\Phi_F}$ and highest $\mathrm {c_p}$: this is reasonable because we have already inferred that the PEG and Ficoll70 do not associate with each other.

%\begin{figure*}[bht]
%	\centering
%	\includegraphics[height=61mm]{Ficoll_aggregate_size.png}
%	\includegraphics[height=61mm]{Fraction_monomer.png}
%	\begin{subfigure} {.4\textwidth} \centering
%\includegraphics[height=61mm]{HydroFunction_Ficoll_combined_.png}%\end{subfigure} %
%	\begin{subfigure} {.4\textwidth} \centering
%	\includegraphics[width=70mm]{HydroFunction_ChargedFicoll_combined_.png}
%	\caption{(a) Fraction of Ficoll monomer and (b) Ficoll aggregate size for both charged and uncharged Ficoll.
%	}
%	\label{fig:DFicollMonomerAggregate}
%\end{figure*}

\section{Discussion and Conclusion}\label{sec:discuss}
In this work, we examine the dynamics of Ficoll70 in water, for both uncharged and charged system. Ficoll70, an often-used artificial crowder, is not hard-sphere-like. This has been indicated elsewhere~\cite{Fissell2007,VenturoliF605}, but our water diffusion measurements suggest that 60\% of the water is surface-associated in the crowding limit, indicating that the polysucrose particle is highly porous. Even more surprisingly, Ficoll70 diffusivity is bi-modal, indicating that it self-clusters at modest concentrations, with cluster sizes approaching 2 to 3 times the size of the single Ficoll70 particle size (``monomer''). This is reminiscent of indications, from maximum entropy analyses of fluorescence correlation spectroscopy experiments, of multiple modes of probe mobility in crowded solutions.~\cite{goins_macromolecular_2008}

Coexistence of monomers and clusters in equilibrium has been seen experimentally~\cite{stradner_equilibrium_2004, barhoum_clusters_2010, porcar_formation_2010}, and is expected in systems which have short-ranged attractions and longer-ranged repulsions.~\cite{groenewold_anomalously_2001, sweatman_cluster_2014} Considering both the 5 nm particle scale and that polysaccharide surfaces in water have a Hamaker constant of $\sim \mathrm {2 k_B T}$~\cite{holmberg_surface_1997}), attractive forces should be relevant in the presence of even small long-ranged (e.g. electrostatic) repulsions, and is consistent with the observed weak clustering.

\begin{acknowledgments}
This work was supported by the Natural Sciences and Engineering Research Council of Canada. 
We thank William Fissell for generously providing us with charged Ficoll70, and Arun Yethiraj, and Francesco Piazza for illuminating discussions.
\end{acknowledgments}

%\nocite{*}
%\bibliography{crowding2}% Produces the bibliography via BibTeX.

\begin{thebibliography}{41}%
\makeatletter
\providecommand \@ifxundefined [1]{%
 \@ifx{#1\undefined}
}%
\providecommand \@ifnum [1]{%
 \ifnum #1\expandafter \@firstoftwo
 \else \expandafter \@secondoftwo
 \fi
}%
\providecommand \@ifx [1]{%
 \ifx #1\expandafter \@firstoftwo
 \else \expandafter \@secondoftwo
 \fi
}%
\providecommand \natexlab [1]{#1}%
\providecommand \enquote  [1]{``#1''}%
\providecommand \bibnamefont  [1]{#1}%
\providecommand \bibfnamefont [1]{#1}%
\providecommand \citenamefont [1]{#1}%
\providecommand \href@noop [0]{\@secondoftwo}%
\providecommand \href [0]{\begingroup \@sanitize@url \@href}%
\providecommand \@href[1]{\@@startlink{#1}\@@href}%
\providecommand \@@href[1]{\endgroup#1\@@endlink}%
\providecommand \@sanitize@url [0]{\catcode `\\12\catcode `\$12\catcode
  `\&12\catcode `\#12\catcode `\^12\catcode `\_12\catcode `\%12\relax}%
\providecommand \@@startlink[1]{}%
\providecommand \@@endlink[0]{}%
\providecommand \url  [0]{\begingroup\@sanitize@url \@url }%
\providecommand \@url [1]{\endgroup\@href {#1}{\urlprefix }}%
\providecommand \urlprefix  [0]{URL }%
\providecommand \Eprint [0]{\href }%
\providecommand \doibase [0]{http://dx.doi.org/}%
\providecommand \selectlanguage [0]{\@gobble}%
\providecommand \bibinfo  [0]{\@secondoftwo}%
\providecommand \bibfield  [0]{\@secondoftwo}%
\providecommand \translation [1]{[#1]}%
\providecommand \BibitemOpen [0]{}%
\providecommand \bibitemStop [0]{}%
\providecommand \bibitemNoStop [0]{.\EOS\space}%
\providecommand \EOS [0]{\spacefactor3000\relax}%
\providecommand \BibitemShut  [1]{\csname bibitem#1\endcsname}%
\let\auto@bib@innerbib\@empty
%</preamble>
\bibitem [{\citenamefont {Palit}\ \emph
  {et~al.}(2017{\natexlab{a}})\citenamefont {Palit}, \citenamefont {He},
  \citenamefont {Hamilton}, \citenamefont {Yethiraj},\ and\ \citenamefont
  {Yethiraj}}]{palit_combining_2017}%
  \BibitemOpen
  \bibfield  {author} {\bibinfo {author} {\bibfnamefont {S.}~\bibnamefont
  {Palit}}, \bibinfo {author} {\bibfnamefont {L.}~\bibnamefont {He}}, \bibinfo
  {author} {\bibfnamefont {W.~A.}\ \bibnamefont {Hamilton}}, \bibinfo {author}
  {\bibfnamefont {A.}~\bibnamefont {Yethiraj}}, \ and\ \bibinfo {author}
  {\bibfnamefont {A.}~\bibnamefont {Yethiraj}},\ }\href@noop {} {\bibfield
  {journal} {\bibinfo  {journal} {Physical Review Letters}\ }\textbf {\bibinfo
  {volume} {118}},\ \bibinfo {pages} {097801} (\bibinfo {year}
  {2017}{\natexlab{a}})}\BibitemShut {NoStop}%
\bibitem [{\citenamefont {Wang}\ \emph
  {et~al.}(2012{\natexlab{a}})\citenamefont {Wang}, \citenamefont {Sarkar},
  \citenamefont {Smith}, \citenamefont {Krois},\ and\ \citenamefont
  {Pielak}}]{wang_macromolecular_2012}%
  \BibitemOpen
  \bibfield  {author} {\bibinfo {author} {\bibfnamefont {Y.}~\bibnamefont
  {Wang}}, \bibinfo {author} {\bibfnamefont {M.}~\bibnamefont {Sarkar}},
  \bibinfo {author} {\bibfnamefont {A.~E.}\ \bibnamefont {Smith}}, \bibinfo
  {author} {\bibfnamefont {A.~S.}\ \bibnamefont {Krois}}, \ and\ \bibinfo
  {author} {\bibfnamefont {G.~J.}\ \bibnamefont {Pielak}},\ }\href@noop {}
  {\bibfield  {journal} {\bibinfo  {journal} {Journal of American Chemical
  Society}\ }\textbf {\bibinfo {volume} {134}},\ \bibinfo {pages} {16614}
  (\bibinfo {year} {2012}{\natexlab{a}})}\BibitemShut {NoStop}%
\bibitem [{\citenamefont {Wong}\ \emph {et~al.}(2016)\citenamefont {Wong},
  \citenamefont {Sandlin}, \citenamefont {Carey}, \citenamefont {Miller},
  \citenamefont {Shank}, \citenamefont {Oklu}, \citenamefont {Maheswaran},
  \citenamefont {Haber}, \citenamefont {Irimia}, \citenamefont {Stott} \emph
  {et~al.}}]{wong2016role}%
  \BibitemOpen
  \bibfield  {author} {\bibinfo {author} {\bibfnamefont {K.~H.}\ \bibnamefont
  {Wong}}, \bibinfo {author} {\bibfnamefont {R.~D.}\ \bibnamefont {Sandlin}},
  \bibinfo {author} {\bibfnamefont {T.~R.}\ \bibnamefont {Carey}}, \bibinfo
  {author} {\bibfnamefont {K.~L.}\ \bibnamefont {Miller}}, \bibinfo {author}
  {\bibfnamefont {A.~T.}\ \bibnamefont {Shank}}, \bibinfo {author}
  {\bibfnamefont {R.}~\bibnamefont {Oklu}}, \bibinfo {author} {\bibfnamefont
  {S.}~\bibnamefont {Maheswaran}}, \bibinfo {author} {\bibfnamefont {D.~A.}\
  \bibnamefont {Haber}}, \bibinfo {author} {\bibfnamefont {D.}~\bibnamefont
  {Irimia}}, \bibinfo {author} {\bibfnamefont {S.~L.}\ \bibnamefont {Stott}},
  \emph {et~al.},\ }\href@noop {} {\bibfield  {journal} {\bibinfo  {journal}
  {Scientific reports}\ }\textbf {\bibinfo {volume} {6}} (\bibinfo {year}
  {2016})}\BibitemShut {NoStop}%
\bibitem [{\citenamefont {Groszek}\ \emph {et~al.}(2010)\citenamefont
  {Groszek}, \citenamefont {Li}, \citenamefont {Ferrell}, \citenamefont
  {Smith}, \citenamefont {Zorman}, \citenamefont {Hofmann}, \citenamefont
  {Roy},\ and\ \citenamefont {Fissell}}]{Groszek_2010}%
  \BibitemOpen
  \bibfield  {author} {\bibinfo {author} {\bibfnamefont {J.}~\bibnamefont
  {Groszek}}, \bibinfo {author} {\bibfnamefont {L.}~\bibnamefont {Li}},
  \bibinfo {author} {\bibfnamefont {N.}~\bibnamefont {Ferrell}}, \bibinfo
  {author} {\bibfnamefont {R.}~\bibnamefont {Smith}}, \bibinfo {author}
  {\bibfnamefont {C.~A.}\ \bibnamefont {Zorman}}, \bibinfo {author}
  {\bibfnamefont {C.~L.}\ \bibnamefont {Hofmann}}, \bibinfo {author}
  {\bibfnamefont {S.}~\bibnamefont {Roy}}, \ and\ \bibinfo {author}
  {\bibfnamefont {W.~H.}\ \bibnamefont {Fissell}},\ }\href@noop {} {\bibfield
  {journal} {\bibinfo  {journal} {American Journal of Physiology - Renal
  Physiology}\ }\textbf {\bibinfo {volume} {299}},\ \bibinfo {pages} {F752}
  (\bibinfo {year} {2010})}\BibitemShut {NoStop}%
\bibitem [{\citenamefont {Fissell}\ \emph {et~al.}(2007)\citenamefont
  {Fissell}, \citenamefont {Manley}, \citenamefont {Dubnisheva}, \citenamefont
  {Glass}, \citenamefont {Magistrelli}, \citenamefont {Eldridge}, \citenamefont
  {Fleischman}, \citenamefont {Zydney},\ and\ \citenamefont
  {Roy}}]{Fissell2007}%
  \BibitemOpen
  \bibfield  {author} {\bibinfo {author} {\bibfnamefont {W.~H.}\ \bibnamefont
  {Fissell}}, \bibinfo {author} {\bibfnamefont {S.}~\bibnamefont {Manley}},
  \bibinfo {author} {\bibfnamefont {A.}~\bibnamefont {Dubnisheva}}, \bibinfo
  {author} {\bibfnamefont {J.}~\bibnamefont {Glass}}, \bibinfo {author}
  {\bibfnamefont {J.}~\bibnamefont {Magistrelli}}, \bibinfo {author}
  {\bibfnamefont {A.~N.}\ \bibnamefont {Eldridge}}, \bibinfo {author}
  {\bibfnamefont {A.~J.}\ \bibnamefont {Fleischman}}, \bibinfo {author}
  {\bibfnamefont {A.~L.}\ \bibnamefont {Zydney}}, \ and\ \bibinfo {author}
  {\bibfnamefont {S.}~\bibnamefont {Roy}},\ }\href {\doibase
  10.1152/ajprenal.00097.2007} {\bibfield  {journal} {\bibinfo  {journal}
  {American Journal of Physiology - Renal Physiology}\ }\textbf {\bibinfo
  {volume} {293}},\ \bibinfo {pages} {F1209} (\bibinfo {year}
  {2007})}\BibitemShut {NoStop}%
\bibitem [{\citenamefont {Venturoli}\ and\ \citenamefont
  {Rippe}(2005)}]{VenturoliF605}%
  \BibitemOpen
  \bibfield  {author} {\bibinfo {author} {\bibfnamefont {D.}~\bibnamefont
  {Venturoli}}\ and\ \bibinfo {author} {\bibfnamefont {B.}~\bibnamefont
  {Rippe}},\ }\href {\doibase 10.1152/ajprenal.00171.2004} {\bibfield
  {journal} {\bibinfo  {journal} {American Journal of Physiology - Renal
  Physiology}\ }\textbf {\bibinfo {volume} {288}},\ \bibinfo {pages} {F605}
  (\bibinfo {year} {2005})}\BibitemShut {NoStop}%
\bibitem [{\citenamefont {Dhar}\ \emph {et~al.}(2010)\citenamefont {Dhar},
  \citenamefont {Samiotakis}, \citenamefont {Ebbinghaus}, \citenamefont
  {Nienhaus}, \citenamefont {Homouz}, \citenamefont {Gruebele},\ and\
  \citenamefont {Cheung}}]{Dhar12102010}%
  \BibitemOpen
  \bibfield  {author} {\bibinfo {author} {\bibfnamefont {A.}~\bibnamefont
  {Dhar}}, \bibinfo {author} {\bibfnamefont {A.}~\bibnamefont {Samiotakis}},
  \bibinfo {author} {\bibfnamefont {S.}~\bibnamefont {Ebbinghaus}}, \bibinfo
  {author} {\bibfnamefont {L.}~\bibnamefont {Nienhaus}}, \bibinfo {author}
  {\bibfnamefont {D.}~\bibnamefont {Homouz}}, \bibinfo {author} {\bibfnamefont
  {M.}~\bibnamefont {Gruebele}}, \ and\ \bibinfo {author} {\bibfnamefont
  {M.~S.}\ \bibnamefont {Cheung}},\ }\href {\doibase 10.1073/pnas.1006760107}
  {\bibfield  {journal} {\bibinfo  {journal} {Proceedings of the National
  Academy of Sciences}\ }\textbf {\bibinfo {volume} {107}},\ \bibinfo {pages}
  {17586} (\bibinfo {year} {2010})}\BibitemShut {NoStop}%
\bibitem [{\citenamefont {Wenner}\ and\ \citenamefont
  {Bloomfield}(1999)}]{Wenner1999}%
  \BibitemOpen
  \bibfield  {author} {\bibinfo {author} {\bibfnamefont {J.~R.}\ \bibnamefont
  {Wenner}}\ and\ \bibinfo {author} {\bibfnamefont {V.~A.}\ \bibnamefont
  {Bloomfield}},\ }\href@noop {} {\bibfield  {journal} {\bibinfo  {journal}
  {Biophysical Journal}\ }\textbf {\bibinfo {volume} {77}},\ \bibinfo {pages}
  {3234–3241} (\bibinfo {year} {1999})}\BibitemShut {NoStop}%
\bibitem [{\citenamefont {Galan}\ \emph {et~al.}(2001)\citenamefont {Galan},
  \citenamefont {Sot}, \citenamefont {Llorca}, \citenamefont {Carrascosa},
  \citenamefont {Valpuesta},\ and\ \citenamefont {Muga}}]{Galan2001}%
  \BibitemOpen
  \bibfield  {author} {\bibinfo {author} {\bibfnamefont {A.}~\bibnamefont
  {Galan}}, \bibinfo {author} {\bibfnamefont {B.}~\bibnamefont {Sot}}, \bibinfo
  {author} {\bibfnamefont {O.}~\bibnamefont {Llorca}}, \bibinfo {author}
  {\bibfnamefont {J.~L.}\ \bibnamefont {Carrascosa}}, \bibinfo {author}
  {\bibfnamefont {J.}~\bibnamefont {Valpuesta}}, \ and\ \bibinfo {author}
  {\bibfnamefont {A.}~\bibnamefont {Muga}},\ }\href@noop {} {\bibfield
  {journal} {\bibinfo  {journal} {Journal of Biological Chemistry}\ }\textbf
  {\bibinfo {volume} {276}},\ \bibinfo {pages} {957} (\bibinfo {year}
  {2001})}\BibitemShut {NoStop}%
\bibitem [{\citenamefont {Lavrenko}, \citenamefont {Mikriukova},\ and\
  \citenamefont {Okatova}(1987)}]{lavrenko1987separation}%
  \BibitemOpen
  \bibfield  {author} {\bibinfo {author} {\bibfnamefont {P.~N.}\ \bibnamefont
  {Lavrenko}}, \bibinfo {author} {\bibfnamefont {O.~I.}\ \bibnamefont
  {Mikriukova}}, \ and\ \bibinfo {author} {\bibfnamefont {O.~V.}\ \bibnamefont
  {Okatova}},\ }\href@noop {} {\bibfield  {journal} {\bibinfo  {journal}
  {Analytical biochemistry}\ }\textbf {\bibinfo {volume} {166}},\ \bibinfo
  {pages} {287} (\bibinfo {year} {1987})}\BibitemShut {NoStop}%
\bibitem [{\citenamefont {Tokuriki}\ \emph {et~al.}(2004)\citenamefont
  {Tokuriki}, \citenamefont {Kinjo}, \citenamefont {Negi}, \citenamefont
  {Hoshino}, \citenamefont {Goto}, \citenamefont {Urabe},\ and\ \citenamefont
  {Yomo}}]{tokuriki2004}%
  \BibitemOpen
  \bibfield  {author} {\bibinfo {author} {\bibfnamefont {N.}~\bibnamefont
  {Tokuriki}}, \bibinfo {author} {\bibfnamefont {M.}~\bibnamefont {Kinjo}},
  \bibinfo {author} {\bibfnamefont {S.}~\bibnamefont {Negi}}, \bibinfo {author}
  {\bibfnamefont {M.}~\bibnamefont {Hoshino}}, \bibinfo {author} {\bibfnamefont
  {Y.}~\bibnamefont {Goto}}, \bibinfo {author} {\bibfnamefont {I.}~\bibnamefont
  {Urabe}}, \ and\ \bibinfo {author} {\bibfnamefont {T.}~\bibnamefont {Yomo}},\
  }\href@noop {} {\bibfield  {journal} {\bibinfo  {journal} {Protein Science}\
  }\textbf {\bibinfo {volume} {13}},\ \bibinfo {pages} {125} (\bibinfo {year}
  {2004})}\BibitemShut {NoStop}%
\bibitem [{\citenamefont {Wang}\ \emph
  {et~al.}(2012{\natexlab{b}})\citenamefont {Wang}, \citenamefont {Benton},
  \citenamefont {Singh},\ and\ \citenamefont {Pielak}}]{wang_disordered_2012}%
  \BibitemOpen
  \bibfield  {author} {\bibinfo {author} {\bibfnamefont {Y.}~\bibnamefont
  {Wang}}, \bibinfo {author} {\bibfnamefont {L.~A.}\ \bibnamefont {Benton}},
  \bibinfo {author} {\bibfnamefont {V.}~\bibnamefont {Singh}}, \ and\ \bibinfo
  {author} {\bibfnamefont {G.~J.}\ \bibnamefont {Pielak}},\ }\href@noop {}
  {\bibfield  {journal} {\bibinfo  {journal} {The Journal of Physical Chemistry
  Letters}\ }\textbf {\bibinfo {volume} {3}},\ \bibinfo {pages} {2703}
  (\bibinfo {year} {2012}{\natexlab{b}})}\BibitemShut {NoStop}%
\bibitem [{\citenamefont {Zimmerman}\ and\ \citenamefont
  {Minton}(1993)}]{zimmerman1993}%
  \BibitemOpen
  \bibfield  {author} {\bibinfo {author} {\bibfnamefont {S.~B.}\ \bibnamefont
  {Zimmerman}}\ and\ \bibinfo {author} {\bibfnamefont {A.~P.}\ \bibnamefont
  {Minton}},\ }\href@noop {} {\bibfield  {journal} {\bibinfo  {journal} {Annual
  Review of Biophysics and Biomolecular Structure}\ }\textbf {\bibinfo {volume}
  {22}},\ \bibinfo {pages} {27} (\bibinfo {year} {1993})}\BibitemShut {NoStop}%
\bibitem [{\citenamefont {Bohrer}, \citenamefont {Patterson},\ and\
  \citenamefont {Carroll}(1984)}]{bohrer1984hindered}%
  \BibitemOpen
  \bibfield  {author} {\bibinfo {author} {\bibfnamefont {M.}~\bibnamefont
  {Bohrer}}, \bibinfo {author} {\bibfnamefont {G.~D.}\ \bibnamefont
  {Patterson}}, \ and\ \bibinfo {author} {\bibfnamefont {P.}~\bibnamefont
  {Carroll}},\ }\href@noop {} {\bibfield  {journal} {\bibinfo  {journal}
  {Macromolecules}\ }\textbf {\bibinfo {volume} {17}},\ \bibinfo {pages} {1170}
  (\bibinfo {year} {1984})}\BibitemShut {NoStop}%
\bibitem [{\citenamefont {Deen}, \citenamefont {Bohrer},\ and\ \citenamefont
  {Epstein}(1981)}]{deen1981effects}%
  \BibitemOpen
  \bibfield  {author} {\bibinfo {author} {\bibfnamefont {W.}~\bibnamefont
  {Deen}}, \bibinfo {author} {\bibfnamefont {M.}~\bibnamefont {Bohrer}}, \ and\
  \bibinfo {author} {\bibfnamefont {N.}~\bibnamefont {Epstein}},\ }\href@noop
  {} {\bibfield  {journal} {\bibinfo  {journal} {AIChE Journal}\ }\textbf
  {\bibinfo {volume} {27}},\ \bibinfo {pages} {952} (\bibinfo {year}
  {1981})}\BibitemShut {NoStop}%
\bibitem [{\citenamefont {Oliver}\ \emph {et~al.}(1992)\citenamefont {Oliver},
  \citenamefont {Anderson}, \citenamefont {Troy}, \citenamefont {Brenner},\
  and\ \citenamefont {Deen}}]{oliver1992determination}%
  \BibitemOpen
  \bibfield  {author} {\bibinfo {author} {\bibfnamefont {J.~D.}\ \bibnamefont
  {Oliver}}, \bibinfo {author} {\bibfnamefont {S.}~\bibnamefont {Anderson}},
  \bibinfo {author} {\bibfnamefont {J.~L.}\ \bibnamefont {Troy}}, \bibinfo
  {author} {\bibfnamefont {B.~M.}\ \bibnamefont {Brenner}}, \ and\ \bibinfo
  {author} {\bibfnamefont {W.}~\bibnamefont {Deen}},\ }\href@noop {} {\bibfield
   {journal} {\bibinfo  {journal} {Journal of the American Society of
  Nephrology}\ }\textbf {\bibinfo {volume} {3}},\ \bibinfo {pages} {214}
  (\bibinfo {year} {1992})}\BibitemShut {NoStop}%
\bibitem [{\citenamefont {Axelsson}\ \emph {et~al.}(2011)\citenamefont
  {Axelsson}, \citenamefont {Sverrisson}, \citenamefont {Rippe}, \citenamefont
  {Fissell},\ and\ \citenamefont {Rippe}}]{axelsson2011reduced}%
  \BibitemOpen
  \bibfield  {author} {\bibinfo {author} {\bibfnamefont {J.}~\bibnamefont
  {Axelsson}}, \bibinfo {author} {\bibfnamefont {K.}~\bibnamefont
  {Sverrisson}}, \bibinfo {author} {\bibfnamefont {A.}~\bibnamefont {Rippe}},
  \bibinfo {author} {\bibfnamefont {W.}~\bibnamefont {Fissell}}, \ and\
  \bibinfo {author} {\bibfnamefont {B.}~\bibnamefont {Rippe}},\ }\href@noop {}
  {\bibfield  {journal} {\bibinfo  {journal} {American Journal of
  Physiology-Renal Physiology}\ }\textbf {\bibinfo {volume} {301}},\ \bibinfo
  {pages} {F708} (\bibinfo {year} {2011})}\BibitemShut {NoStop}%
\bibitem [{\citenamefont {Asgeirsson}\ \emph {et~al.}(2007)\citenamefont
  {Asgeirsson}, \citenamefont {Venturoli}, \citenamefont {Fries}, \citenamefont
  {Rippe},\ and\ \citenamefont {Rippe}}]{asgeirsson2007glomerular}%
  \BibitemOpen
  \bibfield  {author} {\bibinfo {author} {\bibfnamefont {D.}~\bibnamefont
  {Asgeirsson}}, \bibinfo {author} {\bibfnamefont {D.}~\bibnamefont
  {Venturoli}}, \bibinfo {author} {\bibfnamefont {E.}~\bibnamefont {Fries}},
  \bibinfo {author} {\bibfnamefont {B.}~\bibnamefont {Rippe}}, \ and\ \bibinfo
  {author} {\bibfnamefont {C.}~\bibnamefont {Rippe}},\ }\href@noop {}
  {\bibfield  {journal} {\bibinfo  {journal} {Acta Physiologica}\ }\textbf
  {\bibinfo {volume} {191}},\ \bibinfo {pages} {237} (\bibinfo {year}
  {2007})}\BibitemShut {NoStop}%
\bibitem [{\citenamefont {Rippe}\ \emph {et~al.}(2006)\citenamefont {Rippe},
  \citenamefont {Asgeirsson}, \citenamefont {Venturoli}, \citenamefont
  {Rippe},\ and\ \citenamefont {Rippe}}]{Rippe2006}%
  \BibitemOpen
  \bibfield  {author} {\bibinfo {author} {\bibfnamefont {C.}~\bibnamefont
  {Rippe}}, \bibinfo {author} {\bibfnamefont {D.}~\bibnamefont {Asgeirsson}},
  \bibinfo {author} {\bibfnamefont {D.}~\bibnamefont {Venturoli}}, \bibinfo
  {author} {\bibfnamefont {A.}~\bibnamefont {Rippe}}, \ and\ \bibinfo {author}
  {\bibfnamefont {B.}~\bibnamefont {Rippe}},\ }\href@noop {} {\bibfield
  {journal} {\bibinfo  {journal} {Kidney international}\ }\textbf {\bibinfo
  {volume} {69}},\ \bibinfo {pages} {1326} (\bibinfo {year}
  {2006})}\BibitemShut {NoStop}%
\bibitem [{\citenamefont {Ohlson}, \citenamefont {S{\"o}rensson},\ and\
  \citenamefont {Haraldsson}(2000)}]{OhlsonF84}%
  \BibitemOpen
  \bibfield  {author} {\bibinfo {author} {\bibfnamefont {M.}~\bibnamefont
  {Ohlson}}, \bibinfo {author} {\bibfnamefont {J.}~\bibnamefont
  {S{\"o}rensson}}, \ and\ \bibinfo {author} {\bibfnamefont {B.}~\bibnamefont
  {Haraldsson}},\ }\href@noop {} {\bibfield  {journal} {\bibinfo  {journal}
  {American Journal of Physiology - Renal Physiology}\ }\textbf {\bibinfo
  {volume} {279}},\ \bibinfo {pages} {F84} (\bibinfo {year}
  {2000})}\BibitemShut {NoStop}%
\bibitem [{\citenamefont {Asgeirsson}\ \emph {et~al.}(2006)\citenamefont
  {Asgeirsson}, \citenamefont {Venturoli}, \citenamefont {Rippe},\ and\
  \citenamefont {Rippe}}]{asgeirsson2006increased}%
  \BibitemOpen
  \bibfield  {author} {\bibinfo {author} {\bibfnamefont {D.}~\bibnamefont
  {Asgeirsson}}, \bibinfo {author} {\bibfnamefont {D.}~\bibnamefont
  {Venturoli}}, \bibinfo {author} {\bibfnamefont {B.}~\bibnamefont {Rippe}}, \
  and\ \bibinfo {author} {\bibfnamefont {C.}~\bibnamefont {Rippe}},\
  }\href@noop {} {\bibfield  {journal} {\bibinfo  {journal} {American Journal
  of Physiology-Renal Physiology}\ }\textbf {\bibinfo {volume} {291}},\
  \bibinfo {pages} {F1083} (\bibinfo {year} {2006})}\BibitemShut {NoStop}%
\bibitem [{\citenamefont {{\"O}berg}\ and\ \citenamefont
  {Rippe}(2014)}]{oberg2014distributed}%
  \BibitemOpen
  \bibfield  {author} {\bibinfo {author} {\bibfnamefont {C.~M.}\ \bibnamefont
  {{\"O}berg}}\ and\ \bibinfo {author} {\bibfnamefont {B.}~\bibnamefont
  {Rippe}},\ }\href@noop {} {\bibfield  {journal} {\bibinfo  {journal}
  {American Journal of Physiology-Renal Physiology}\ }\textbf {\bibinfo
  {volume} {306}},\ \bibinfo {pages} {F844} (\bibinfo {year}
  {2014})}\BibitemShut {NoStop}%
\bibitem [{\citenamefont {Fissell}\ \emph {et~al.}(2010)\citenamefont
  {Fissell}, \citenamefont {Hofmann}, \citenamefont {Smith},\ and\
  \citenamefont {Chen}}]{fissell2010size}%
  \BibitemOpen
  \bibfield  {author} {\bibinfo {author} {\bibfnamefont {W.~H.}\ \bibnamefont
  {Fissell}}, \bibinfo {author} {\bibfnamefont {C.~L.}\ \bibnamefont
  {Hofmann}}, \bibinfo {author} {\bibfnamefont {R.}~\bibnamefont {Smith}}, \
  and\ \bibinfo {author} {\bibfnamefont {M.~H.}\ \bibnamefont {Chen}},\
  }\href@noop {} {\bibfield  {journal} {\bibinfo  {journal} {American Journal
  of Physiology-Renal Physiology}\ }\textbf {\bibinfo {volume} {298}},\
  \bibinfo {pages} {F205} (\bibinfo {year} {2010})}\BibitemShut {NoStop}%
\bibitem [{\citenamefont {Lavrenko}, \citenamefont {Mikryukova},\ and\
  \citenamefont {Didenko}(1986)}]{lavrenko1986hydrodynamic}%
  \BibitemOpen
  \bibfield  {author} {\bibinfo {author} {\bibfnamefont {P.}~\bibnamefont
  {Lavrenko}}, \bibinfo {author} {\bibfnamefont {O.}~\bibnamefont
  {Mikryukova}}, \ and\ \bibinfo {author} {\bibfnamefont {S.}~\bibnamefont
  {Didenko}},\ }\href@noop {} {\bibfield  {journal} {\bibinfo  {journal}
  {Polymer Science USSR}\ }\textbf {\bibinfo {volume} {28}},\ \bibinfo {pages}
  {576} (\bibinfo {year} {1986})}\BibitemShut {NoStop}%
\bibitem [{\citenamefont {Georgalis}\ \emph {et~al.}(2012)\citenamefont
  {Georgalis}, \citenamefont {Philipp}, \citenamefont {Aleksandrova},\ and\
  \citenamefont {Kr{\"u}ger}}]{georgalis2012light}%
  \BibitemOpen
  \bibfield  {author} {\bibinfo {author} {\bibfnamefont {Y.}~\bibnamefont
  {Georgalis}}, \bibinfo {author} {\bibfnamefont {M.}~\bibnamefont {Philipp}},
  \bibinfo {author} {\bibfnamefont {R.}~\bibnamefont {Aleksandrova}}, \ and\
  \bibinfo {author} {\bibfnamefont {J.}~\bibnamefont {Kr{\"u}ger}},\
  }\href@noop {} {\bibfield  {journal} {\bibinfo  {journal} {Journal of colloid
  and interface science}\ }\textbf {\bibinfo {volume} {386}},\ \bibinfo {pages}
  {141} (\bibinfo {year} {2012})}\BibitemShut {NoStop}%
\bibitem [{\citenamefont {Palit}\ \emph
  {et~al.}(2017{\natexlab{b}})\citenamefont {Palit}, \citenamefont {He},
  \citenamefont {Hamilton}, \citenamefont {Yethiraj},\ and\ \citenamefont
  {Yethiraj}}]{sp2}%
  \BibitemOpen
  \bibfield  {author} {\bibinfo {author} {\bibfnamefont {S.}~\bibnamefont
  {Palit}}, \bibinfo {author} {\bibfnamefont {L.}~\bibnamefont {He}}, \bibinfo
  {author} {\bibfnamefont {W.~A.}\ \bibnamefont {Hamilton}}, \bibinfo {author}
  {\bibfnamefont {A.}~\bibnamefont {Yethiraj}}, \ and\ \bibinfo {author}
  {\bibfnamefont {A.}~\bibnamefont {Yethiraj}},\ }\href@noop {} {\bibfield
  {journal} {\bibinfo  {journal} {Journal of Chemical Physics}\ ,\ \bibinfo
  {pages} {submitted}} (\bibinfo {year} {2017}{\natexlab{b}})}\BibitemShut
  {NoStop}%
\bibitem [{\citenamefont {Groenewold}\ and\ \citenamefont
  {Kegel}(2001)}]{groenewold_anomalously_2001}%
  \BibitemOpen
  \bibfield  {author} {\bibinfo {author} {\bibfnamefont {J.}~\bibnamefont
  {Groenewold}}\ and\ \bibinfo {author} {\bibfnamefont {W.~K.}\ \bibnamefont
  {Kegel}},\ }\href@noop {} {\bibfield  {journal} {\bibinfo  {journal} {The
  Journal of Physical Chemistry B}\ }\textbf {\bibinfo {volume} {105}},\
  \bibinfo {pages} {11702} (\bibinfo {year} {2001})}\BibitemShut {NoStop}%
\bibitem [{\citenamefont {Stradner}\ \emph {et~al.}(2004)\citenamefont
  {Stradner}, \citenamefont {Sedgwick}, \citenamefont {Cardinaux},
  \citenamefont {Poon}, \citenamefont {Egelhaaf},\ and\ \citenamefont
  {Schurtenberger}}]{stradner_equilibrium_2004}%
  \BibitemOpen
  \bibfield  {author} {\bibinfo {author} {\bibfnamefont {A.}~\bibnamefont
  {Stradner}}, \bibinfo {author} {\bibfnamefont {H.}~\bibnamefont {Sedgwick}},
  \bibinfo {author} {\bibfnamefont {F.}~\bibnamefont {Cardinaux}}, \bibinfo
  {author} {\bibfnamefont {W.}~\bibnamefont {Poon}}, \bibinfo {author}
  {\bibfnamefont {S.}~\bibnamefont {Egelhaaf}}, \ and\ \bibinfo {author}
  {\bibfnamefont {P.}~\bibnamefont {Schurtenberger}},\ }\href {\doibase
  10.1038/nature03109} {\bibfield  {journal} {\bibinfo  {journal} {Nature}\
  }\textbf {\bibinfo {volume} {432}},\ \bibinfo {pages} {492} (\bibinfo {year}
  {2004})}\BibitemShut {NoStop}%
\bibitem [{\citenamefont {Barhoum}\ and\ \citenamefont
  {Yethiraj}(2010)}]{barhoum_clusters_2010}%
  \BibitemOpen
  \bibfield  {author} {\bibinfo {author} {\bibfnamefont {S.}~\bibnamefont
  {Barhoum}}\ and\ \bibinfo {author} {\bibfnamefont {A.}~\bibnamefont
  {Yethiraj}},\ }\href@noop {} {\bibfield  {journal} {\bibinfo  {journal} {The
  Journal of Physical Chemistry B}\ }\textbf {\bibinfo {volume} {114}},\
  \bibinfo {pages} {17062} (\bibinfo {year} {2010})}\BibitemShut {NoStop}%
\bibitem [{\citenamefont {Porcar}\ \emph {et~al.}(2010)\citenamefont {Porcar},
  \citenamefont {Falus}, \citenamefont {Chen}, \citenamefont {Faraone},
  \citenamefont {Fratini}, \citenamefont {Hong}, \citenamefont {Baglioni},\
  and\ \citenamefont {Liu}}]{porcar_formation_2010}%
  \BibitemOpen
  \bibfield  {author} {\bibinfo {author} {\bibfnamefont {L.}~\bibnamefont
  {Porcar}}, \bibinfo {author} {\bibfnamefont {P.}~\bibnamefont {Falus}},
  \bibinfo {author} {\bibfnamefont {W.-R.}\ \bibnamefont {Chen}}, \bibinfo
  {author} {\bibfnamefont {A.}~\bibnamefont {Faraone}}, \bibinfo {author}
  {\bibfnamefont {E.}~\bibnamefont {Fratini}}, \bibinfo {author} {\bibfnamefont
  {K.}~\bibnamefont {Hong}}, \bibinfo {author} {\bibfnamefont {P.}~\bibnamefont
  {Baglioni}}, \ and\ \bibinfo {author} {\bibfnamefont {Y.}~\bibnamefont
  {Liu}},\ }\href@noop {} {\bibfield  {journal} {\bibinfo  {journal} {The
  Journal of Physical Chemistry Letters}\ }\textbf {\bibinfo {volume} {1}},\
  \bibinfo {pages} {126} (\bibinfo {year} {2010})}\BibitemShut {NoStop}%
\bibitem [{\citenamefont {Barhoum}, \citenamefont {Agarwal},\ and\
  \citenamefont {Yethiraj}(2013)}]{barhoum_cluster_2013}%
  \BibitemOpen
  \bibfield  {author} {\bibinfo {author} {\bibfnamefont {S.}~\bibnamefont
  {Barhoum}}, \bibinfo {author} {\bibfnamefont {A.}~\bibnamefont {Agarwal}}, \
  and\ \bibinfo {author} {\bibfnamefont {A.}~\bibnamefont {Yethiraj}},\ }in\
  \href@noop {} {\emph {\bibinfo {booktitle} {New {Challenges} in
  {Electrostatics} of {Soft} and {Disordered} {Matter}}}},\ \bibinfo {editor}
  {edited by\ \bibinfo {editor} {\bibfnamefont {D.}~\bibnamefont {Dean}},
  \bibinfo {editor} {\bibfnamefont {J.}~\bibnamefont {Dobnikar}}, \bibinfo
  {editor} {\bibfnamefont {A.}~\bibnamefont {Naji}}, \ and\ \bibinfo {editor}
  {\bibfnamefont {R.}~\bibnamefont {Podgornik}}}\ (\bibinfo  {publisher} {Pan
  Stanford},\ \bibinfo {year} {2013})\BibitemShut {NoStop}%
\bibitem [{\citenamefont {Sweatman}, \citenamefont {Fartaria},\ and\
  \citenamefont {Lue}(2014)}]{sweatman_cluster_2014}%
  \BibitemOpen
  \bibfield  {author} {\bibinfo {author} {\bibfnamefont {M.~B.}\ \bibnamefont
  {Sweatman}}, \bibinfo {author} {\bibfnamefont {R.}~\bibnamefont {Fartaria}},
  \ and\ \bibinfo {author} {\bibfnamefont {L.}~\bibnamefont {Lue}},\
  }\href@noop {} {\bibfield  {journal} {\bibinfo  {journal} {The Journal of
  Chemical Physics}\ }\textbf {\bibinfo {volume} {140}},\ \bibinfo {pages}
  {124508} (\bibinfo {year} {2014})}\BibitemShut {NoStop}%
\bibitem [{\citenamefont {S{\"o}rensson}\ \emph {et~al.}(1998)\citenamefont
  {S{\"o}rensson}, \citenamefont {Ohlson}, \citenamefont {Lindstr{\"o}m},\ and\
  \citenamefont {Haraldsson}}]{sorensson1998glomerular}%
  \BibitemOpen
  \bibfield  {author} {\bibinfo {author} {\bibfnamefont {J.}~\bibnamefont
  {S{\"o}rensson}}, \bibinfo {author} {\bibfnamefont {M.}~\bibnamefont
  {Ohlson}}, \bibinfo {author} {\bibfnamefont {K.}~\bibnamefont
  {Lindstr{\"o}m}}, \ and\ \bibinfo {author} {\bibfnamefont {B.}~\bibnamefont
  {Haraldsson}},\ }\href@noop {} {\bibfield  {journal} {\bibinfo  {journal}
  {Acta physiologica Scandinavica}\ }\textbf {\bibinfo {volume} {163}},\
  \bibinfo {pages} {83} (\bibinfo {year} {1998})}\BibitemShut {NoStop}%
\bibitem [{\citenamefont {Luby-Phelps}\ \emph {et~al.}(1987)\citenamefont
  {Luby-Phelps}, \citenamefont {Castle}, \citenamefont {Taylor},\ and\
  \citenamefont {Lanni}}]{luby1987hindered}%
  \BibitemOpen
  \bibfield  {author} {\bibinfo {author} {\bibfnamefont {K.}~\bibnamefont
  {Luby-Phelps}}, \bibinfo {author} {\bibfnamefont {P.~E.}\ \bibnamefont
  {Castle}}, \bibinfo {author} {\bibfnamefont {D.~L.}\ \bibnamefont {Taylor}},
  \ and\ \bibinfo {author} {\bibfnamefont {F.}~\bibnamefont {Lanni}},\
  }\href@noop {} {\bibfield  {journal} {\bibinfo  {journal} {Proceedings of the
  National Academy of Sciences}\ }\textbf {\bibinfo {volume} {84}},\ \bibinfo
  {pages} {4910} (\bibinfo {year} {1987})}\BibitemShut {NoStop}%
\bibitem [{\citenamefont {Price}(1997)}]{Price1997}%
  \BibitemOpen
  \bibfield  {author} {\bibinfo {author} {\bibfnamefont {W.~S.}\ \bibnamefont
  {Price}},\ }\href@noop {} {\bibfield  {journal} {\bibinfo  {journal}
  {Concepts in Magnetic Resonance}\ }\textbf {\bibinfo {volume} {9}},\ \bibinfo
  {pages} {299} (\bibinfo {year} {1997})}\BibitemShut {NoStop}%
\bibitem [{\citenamefont {Barhoum}, \citenamefont {Palit},\ and\ \citenamefont
  {Yethiraj}(2016)}]{barhoum_diffusion_2016}%
  \BibitemOpen
  \bibfield  {author} {\bibinfo {author} {\bibfnamefont {S.}~\bibnamefont
  {Barhoum}}, \bibinfo {author} {\bibfnamefont {S.}~\bibnamefont {Palit}}, \
  and\ \bibinfo {author} {\bibfnamefont {A.}~\bibnamefont {Yethiraj}},\
  }\href@noop {} {\bibfield  {journal} {\bibinfo  {journal} {Progress in
  Nuclear Magnetic Resonance Spectroscopy}\ }\textbf {\bibinfo {volume}
  {94-95}},\ \bibinfo {pages} {1} (\bibinfo {year} {2016})}\BibitemShut
  {NoStop}%
\bibitem [{\citenamefont {Rosenfeld}(1977)}]{rosenfeld_relation_1977}%
  \BibitemOpen
  \bibfield  {author} {\bibinfo {author} {\bibfnamefont {Y.}~\bibnamefont
  {Rosenfeld}},\ }\href@noop {} {\bibfield  {journal} {\bibinfo  {journal}
  {Physical Review A}\ }\textbf {\bibinfo {volume} {15}},\ \bibinfo {pages}
  {2545} (\bibinfo {year} {1977})}\BibitemShut {NoStop}%
\bibitem [{\citenamefont {Dzugutov}(1996)}]{dzugutov_universal_1996}%
  \BibitemOpen
  \bibfield  {author} {\bibinfo {author} {\bibfnamefont {M.}~\bibnamefont
  {Dzugutov}},\ }\href@noop {} {\bibfield  {journal} {\bibinfo  {journal}
  {Nature}\ }\textbf {\bibinfo {volume} {381}},\ \bibinfo {pages} {137}
  (\bibinfo {year} {1996})}\BibitemShut {NoStop}%
\bibitem [{\citenamefont {Thorneywork}\ \emph {et~al.}(2015)\citenamefont
  {Thorneywork}, \citenamefont {Rozas}, \citenamefont {Dullens},\ and\
  \citenamefont {Horbach}}]{thorneywork_effect_2015}%
  \BibitemOpen
  \bibfield  {author} {\bibinfo {author} {\bibfnamefont {A.~L.}\ \bibnamefont
  {Thorneywork}}, \bibinfo {author} {\bibfnamefont {R.~E.}\ \bibnamefont
  {Rozas}}, \bibinfo {author} {\bibfnamefont {R.~P.}\ \bibnamefont {Dullens}},
  \ and\ \bibinfo {author} {\bibfnamefont {J.}~\bibnamefont {Horbach}},\
  }\href@noop {} {\bibfield  {journal} {\bibinfo  {journal} {Physical Review
  Letters}\ }\textbf {\bibinfo {volume} {115}},\ \bibinfo {pages} {268301}
  (\bibinfo {year} {2015})}\BibitemShut {NoStop}%
\bibitem [{\citenamefont {Goins}, \citenamefont {Sanabria},\ and\ \citenamefont
  {Waxham}(2008)}]{goins_macromolecular_2008}%
  \BibitemOpen
  \bibfield  {author} {\bibinfo {author} {\bibfnamefont {A.~B.}\ \bibnamefont
  {Goins}}, \bibinfo {author} {\bibfnamefont {H.}~\bibnamefont {Sanabria}}, \
  and\ \bibinfo {author} {\bibfnamefont {M.~N.}\ \bibnamefont {Waxham}},\
  }\href@noop {} {\bibfield  {journal} {\bibinfo  {journal} {Biophysical
  Journal}\ }\textbf {\bibinfo {volume} {95}},\ \bibinfo {pages} {5362}
  (\bibinfo {year} {2008})}\BibitemShut {NoStop}%
\bibitem [{\citenamefont {Holmberg}\ \emph {et~al.}(1997)\citenamefont
  {Holmberg}, \citenamefont {Berg}, \citenamefont {Stemme}, \citenamefont
  {\"Odberg}, \citenamefont {Rasmusson},\ and\ \citenamefont
  {Claesson}}]{holmberg_surface_1997}%
  \BibitemOpen
  \bibfield  {author} {\bibinfo {author} {\bibfnamefont {M.}~\bibnamefont
  {Holmberg}}, \bibinfo {author} {\bibfnamefont {J.}~\bibnamefont {Berg}},
  \bibinfo {author} {\bibfnamefont {S.}~\bibnamefont {Stemme}}, \bibinfo
  {author} {\bibfnamefont {L.}~\bibnamefont {\"Odberg}}, \bibinfo {author}
  {\bibfnamefont {J.}~\bibnamefont {Rasmusson}}, \ and\ \bibinfo {author}
  {\bibfnamefont {P.}~\bibnamefont {Claesson}},\ }\href@noop {} {\bibfield
  {journal} {\bibinfo  {journal} {Journal of Colloid and Interface Science}\
  }\textbf {\bibinfo {volume} {186}},\ \bibinfo {pages} {369} (\bibinfo {year}
  {1997})}\BibitemShut {NoStop}%
\end{thebibliography}

\end{document}